\documentclass[useAMS,usenatbib,usegraphicx]{mn2e}
\newcommand{\Ox}{[\mbox{O\,{\sc iii}}]}
\newcommand{\C}{\mbox{C\,{\sc iv}}}

\title[On the origin of radio emission in radio quiet quasars]{On the origin of radio emission in radio quiet quasars}
\author[A. Laor and E. Behar]
{Ari Laor$^1$\thanks{E-mail: laor@physics.technion.ac.il (AL); behar@physics.technion.ac.il (EB)} and 
Ehud Behar$^{1,2}$
\footnotemark[1] \\ 
$^1$Physics Department, Technion, Haifa~32000, Israel \\
$^2$Current address: Senior NPP Fellow, NASA / Goddard Space Flight Center, Greenbelt MD 20771}

\begin{document}
\maketitle

%\date{Accepted 1988 December 15. Received 1988 December 14; in original form 1988 October 11}

\pagerange{\pageref{firstpage}--\pageref{lastpage}} \pubyear{2002}

\maketitle

\label{firstpage}

\begin{abstract}
The radio emission in radio loud quasars originates in a jet carrying relativistic
electrons. In radio quiet quasars (RQQs) the relative radio emission is $\sim 10^3$ times 
weaker, and its origin is not established yet. We show here that 
there is a strong correlation between the radio luminosity ($L_R$) and X-ray luminosity 
($L_X$) with $L_R\sim 10^{-5}L_X$, for the radio quiet Palomar-Green (PG) quasar sample.
The sample is optically selected, with nearly complete radio and X-ray detections, and thus
this correlation cannot be due to direct selection biases. The PG quasars lie on an extension
of a similar correlation noted by Panessa et al., for a small sample of nearby low luminosity 
type 1 AGN. A remarkably similar correlation, known as the G{\"u}del-Benz relation, where 
$L_R/L_X\sim 10^{-5}$, holds for coronally active stars. The G{\"u}del-Benz relation, 
together with correlated stellar X-ray and radio variability, implies that the coronae are magnetically heated. We therefore raise the possibility that AGN coronae are also magnetically heated, and that 
the radio emission in RQQ also originates in coronal activity. 
If correct, then RQQ should generally display compact flat cores at a few GHz 
due to synchrotron self-absorption, while at a few hundred GHz we should be able
to see directly the X-ray emitting corona, and relatively rapid and large amplitude variability,
correlated with the X-ray variability, is likely to be seen. We also discuss possible evidence that the radio and X-ray emission in ultra luminous X-ray sources and Galactic black 
holes may be of coronal origin as well.

\end{abstract}


\section{Introduction}
AGN emit continuum radiation from the radio to the hard X-rays, and in some objects beyond.
The overall spectral energy distribution (SED) of AGN has a characterstic shape, with 
a relatively small dispersion (e.g. Sanders et al. 1989, Elvis et al. 1994), except in 
the radio band, where the relative strength of the radio emission spans a range of 
$>10^6$, from the most radio loud AGN to the most radio quiet AGN,
where the radio emission is undetectable. Furthermore, there is evidence that the 
distributions of absolute radio power (Miller et al. 1990) and
relative radio power, commonly measured using $R\equiv f_{\rm 6~cm}/f_{\rm 4400~\AA}$,
are bimodal (Kellerman et al. 1989), with radio loud (RL) AGN having 
$R\sim 10^2-10^5$ and radio quiet (RQ) AGN 
having $R\la 0.1-10$, though the strength of the bi-modality 
may be weaker than initially estimated (White et al. 2007). The radio emission of RL
AGN originates in a fast jet carrying relativistic electrons, as clearly established 
through high resolution radio imaging (e.g. Begelman et al. 1984, and citations 
thereafter), however the origin of the radio emission in RQ AGN is not established yet. 

Early radio imaging of RQ AGN confined the radio emission to arcsec scale, which 
allowed the option of supernovae related radio emission from compact starburst
activity (Wilson \& Willis 1980; Ulvestad \& Wilson 1984; Sopp \& Alexander 1991; 
Condon et al. 1991; Terlevich et al. 1992). 
However, sub arcsecond imaging of RQ AGN (Kellerman et al. 1994; 
Kukula et al. 1995, 1998; Leipski et al. 2006),
followed by higher resolution VLBI imaging (Blundell et al. 1996, Blundell \& Beasley 1998; 
Ulvestad \& Ho 2001; Nagar et al. 2002; Anderson et al. 2004; Ulvestad et al. 2005), 
found significant compact radio emission on mas, i.e. pc scale. 
Most recently, radio variability monitoring
of RQ AGN confine the radio emission to $\la 0.1$~pc in luminous AGN 
(Barvainis et al. 2005), and $\la 10^{-4} - 10^{-2}$~pc in low luminosity AGN 
(Anderson \& Ulvestad 2005), which clearly indicates a compact
non thermal source. Unlike high resolution imaging of RL AGN, which often resolves
a significant fraction of the total radio flux already on arcsec (i.e. kpc) 
scale, in RQ AGN the emission is mostly unresolved at arcsec scale, and often
remains largely unresolved down to mas. {\em What produces this 
radio emission?}. A plausible explanation is a scaled down version of the RL AGN
mechanism, i.e. a low power jet (Miller et al. 1993; Falcke et al. 1996), 
which dissipates before leaving the core, explaining
the general lack of resolved emission. However, the bimodal distribution in jet 
power remains to be understood.

Further hints on the origin of the radio emission in RQ AGN can be provided by 
correlating it with other emission properties. In this paper we focus on the 
correlation of the radio emission with the X-ray emission (herafter the R-X relation). 
Earlier studies of this relation  
(Brinkmann et al. 2000; Salvato et al. 
2004; Wang et al. 2006), indicated a roughly linear relation for luminous 
RQ AGN. These studies were based 
on a correlation of the ROSAT all sky survey, and large area radio surveys 
(VLA FIRST and NVSS). Most 
objects in flux limited surveys tend to cluster close to the flux limit, and
thus a linear relation would appear if the range in luminosities is significantly larger
than the range in fluxes, as the luminosity in both bands is then necessarily
correlated with the redshift. More recently, Panessa et al. (2007) found a 
linear R-X relation in an optically selected sample of nearby low luminosity
AGN (see also Terashima \& Wilson 2003). This sample is still somewhat affected
by selection effects and incompleteness, and a significant fraction of the objects 
there may be affected by X-ray absorption, which cannot be accurately corrected.

To avoid a correlation induced by selection effects, one needs a well defined
and complete sample of AGN, which is selected independently of the radio and X-ray 
emission properties, and yet includes complete radio and X-ray detections. 
The best sample for this purpose is PG Bright
Quasar Survey sample (Schmidt \& Green 1983). This sample is selected based on flux 
($B<16.16$),
morphology (point like on the Palomar Schmidt Survey plates), and color ($U-B<-0.44$).
It includes the 114 brightest AGN in $1/4$ of the sky (at high celestial and
Galactic latitudes).
The completeness of the sample has been debated, and the recent thorough study
of Jester et al. (2005) finds a somewhat different effective cut ($U-B<-0.7$), but 
with no systematic incompleteness (at $z<0.5$). The PG quasars are representative
of $B$ band selected quasars, and are generally similar to bright quasars selected
in other wavelengths. The bright magnitude limit facilitates studies of this sample in
many other bands, including the X-ray and radio bands. An unpublished study 
(Ian George, private communication), based on {\em ASCA} observations of 26 PG quasars 
(George et al. 2000), indicated a strong R-X relation. Motivated by this result, 
and the presence of a similarly strong R-X relation in coronally active stars,
we study here the R-X relation in a complete sample of 87 $z\le 0.5$ PG quasars
(Boroson \& Green 1992).
We find that both coronally active stars and active galaxies appear to follow
the same relation, as further discussed below. In section 2 we present the 
R-X relation for the PG quasars, and expand the luminosity range to include
a smaller sample of low luminosity AGN, ultraluminous X-ray sources,
Galactic Black Holes (GBHs), and coronally active stars. In section
3 we discuss various implications for the origin of the radio emission in RQ AGN,
and possible future tests of the suggested origin.

\begin{figure*}
\includegraphics[width=165mm,angle=0]{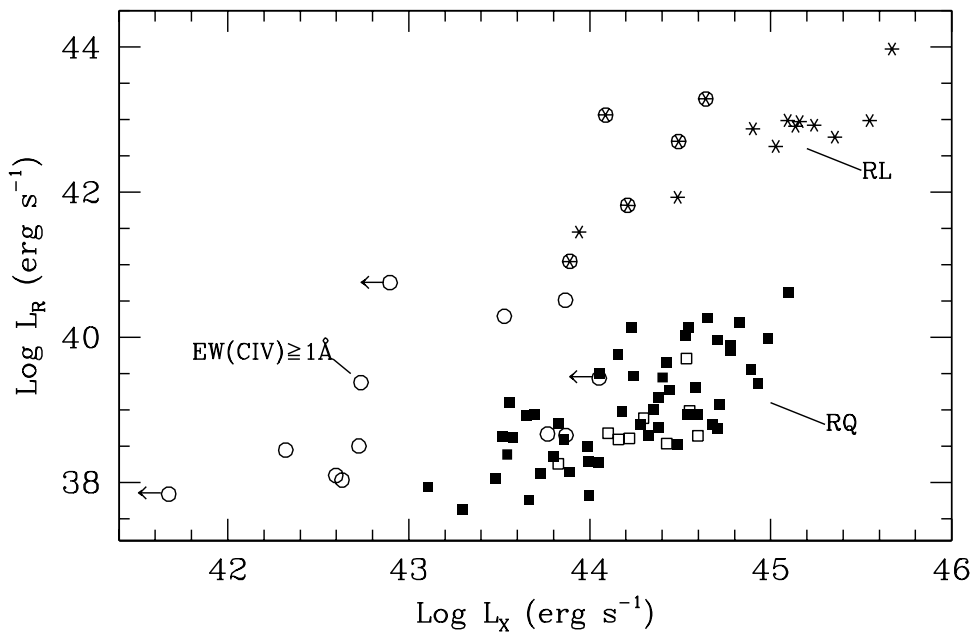}
\caption{The distribution of the PG quasars in the $L_{\rm R}$ vs. $L_{\rm X}$ plane.
RL AGN are designated by stars, and their $L_{\rm R}/L_{\rm X}$ is about a factor of
$10^3$ higher than for RQ AGN. Circles mark objects with \C\ absorption EW$>1$\AA\ (some of
which are RL), where left attached arrows mark upper limits to $L_{\rm X}$. These UV
absorbed AGN tend to be underluminous by a factor  of 10-30 in $L_{\rm X}$
for a given $L_{\rm R}$ (and also for a given optical luminosity), most likely due to
absorption. Squares are RQ
AGN without \C\ absorption, where the filled/empty squares mark objects with radio
detections/upper limits. Note the small scatter in the $L_{\rm R}$ vs. $L_{\rm X}$
correlation in {\em optically} selected AGN, once RL and UV absorbed AGN are excluded.}
\end{figure*}

\section{The Radio - X-ray relation}

\subsection{The PG Quasar Sample}

As noted above, we study the R-X relation in AGN using the optically selected 
PG quasar sample. We use the Boroson \& Green (1992, hereafter BG92) subsample of 
87 PG quasars with $z\le 0.5$, which was studied extensively over a wide range
of energies. The radio fluxes are taken from the relatively deep radio observations
by Kellerman et al. (1989, 1994) with the Very Large Array (VLA), which detected 
$\sim 90$\% (78/87) of the objects in the sample. The X-ray fluxes are taken from 
Brandt, Laor \& Wills (2000) and Laor \& Brandt (2002), who provide detections 
for $\sim 97$\% (84/87) of the objects based on {\em Rosat} observations. 
Table 1 presents all the 
data used for the current analysis. Specifically, column (2) lists the redshifts 
determined using the peak of the \Ox\ $\lambda$5007 line (T. Boroson, private
communication). Column (3) lists M$_{\rm V}$ taken from BG92. Column (4) lists 
log $R$, taken from Kellerman et al. 
(1989, 1994). Column (5) lists the spectral slope calculated between rest frame 
3000~\AA\ and 2~keV, $\alpha_{\rm ox}$, taken from 
Brandt et al. (2000) and Laor \& Brandt (2002). Column (6) lists the observed
flux density at rest frame  1~keV. Columns (7) and (8) list $\nu L_{\nu}$ at
observed 6~cm, L$_{\rm R}$, and the integrated X-ray luminosity over rest frame
0.2-20~keV, L$_{\rm X}$. Finally, column (9)
lists the \C\ line absorption equivalent width, taken from Laor \& Brandt (2002).

We integrate over 0.2-20~keV in order to get an estimate of the bolometric
accretion disk coronal emission. The typical AGN emission rises steeply below 
0.2~keV, most likely due to photospheric accretion disk emission, while above
$\sim 20$~keV the spectrum becomes steeper, as indicated by the shape of the
X-ray background (e.g. Gilli et al. 2007). Thus, the 0.2-20~keV should represent
the bulk of the coronal emission. However, rather than perform a direct integration
for each object, we use the following surprisingly accurate approximation, 
L$_{\rm X}=C\nu$L$_{\nu}$(1~keV) with $C=6.25$. The value
of $C$ is obtained through the following consideration. According to Brocksopp
et al. (2006), the 0.3-10~keV spectra of most PG quasars is well described by a 
broken power-law with a break energy around 1.5~keV (Fig.2 there). 
The soft (below the break) spectral slopes range from $-2.2$ to $-1.35$, and
the hard (above the break) spectral slopes range from $-1.4$ to $-0.85$ .
Furthermore, there is a remarkably tight correlation between the soft and hard 
spectral slopes (correlation coefficient 0.97, Fig.4 there). Integrating
the flattest X-ray spectrum from 0.2 to 20 keV ($-1.35/-0.85$ soft/hard)
we find that $C=5.9$, while for the steepest spectrum ($-2.2/-1.4$ soft/hard)
we get $C=6.6$. Thus, assuming a fixed $C=6.25$ for all objects
leads to an error of at most $\sim 5$\% in the bolometric conversion from 
$\nu$L$_{\nu}$(1~keV) to L$_{\rm X}$, which is negligible for our purpose here.

Figure 1 presents the distribution of the 87 BG92 objects in the L$_{\rm R}$ vs. 
L$_{\rm X}$ plane. There are 71 radio quiet and 16 radio loud AGN in this sample, where we use the
Kellerman et al. (1989) definition for radio loud AGN as having observed
frame $R>10$. As seen in Fig.1, radio loud AGN typically have a $10^3$ 
higher L$_{\rm R}$ compared to radio quiet AGN, at a given L$_{\rm X}$ 
(see also Terashima \& Wilson 2003; Capetti \& Balmaverde 2007).
The origin of the radio emission in radio loud AGN is well established, and we therefore 
do not consider these objects further here. There is strong evidence that X-ray and UV 
absorption occur simultaneously in AGN (Crenshaw et al. 1999; Brandt et al. 2000).
In some specific cases, there is also kinematic similarity between the UV and X-ray absorbers
(Gabel et al. 2003).
We therefore also exclude from our sample all 17 objects having \C\ line absorption equivalent 
width $>1$~\AA\ (Laor \& Brandt 2002\footnote{The Laor \& Brandt measurements are based on high S/N
UV spectra available for only 56 of the 87 BG92 objects. 
However, a more comprehensive
study of the \C\ line profile of 81 of the 87 BG92 objects, including lower quality UV spectra
(Baskin \& Laor 2005) indicates there are no other objects with strong \C\ 
absorption}), of which 12 are radio quiet. This leaves us with a sample of 59 RQQ which are most likely 
unaffected by X-ray absorption. Radio detections are available for 50 of these 59 objects,
and the distribution of the remaining 9 objects in the L$_{\rm R}$ vs. L$_{\rm X}$ plane
is consistent with the distribution of the 50 objects with X-ray and radio detections. 
As clearly 
seen in Fig.1, the exclusion of RLQ and of quasars with \C\ absorption reduces significantly 
the scatter of the optically selected quasars in the L$_{\rm R}$ vs. L$_{\rm X}$ plane.

Figure 2a presents L$_{\rm R}$ vs. L$_{\rm X}$ for the 59 unabsorbed RQQ together with the
best fit linear correlations. Minimizing the scatter in L$_{\rm R}$ ($L_X$ being the independent variable)
gives a best fit linear relation 
\begin{equation}
\log L_{\rm R, 39}=-0.21\pm 0.08+(1.08\pm 0.15)\log L_{\rm X, 44} 
\end{equation}
with a dispersion of $\sigma=0.51$ in log~$L_{\rm R}$, where $L_{\rm R, 39}=L_{\rm R}/10^{39}$~erg~s$^{-1}$,
and $L_{\rm X, 44}=L_{\rm X}/10^{44}$~erg~s$^{-1}$. Minimizing the scatter in L$_{\rm X}$ 
($L_R$ independent) gives 
\begin{equation}
\log L_{\rm X, 44}=0.21\pm 0.05+(0.48\pm 0.06)\log L_{\rm R, 39} 
\end{equation}
with a dispersion of $\sigma=0.34$ in log~$L_{\rm X}$. The Spearman rank order correlation
coefficient is $R_S=0.71$, which has a chance probability of $P_r=6.5\times 10^{-9}$. The X-ray flux is correlated with the radio flux with $R_S=0.50$ and $P_r=2\times 10^{-5}$, which
indicates that the correlation of $L_{\rm X}$ and $L_{\rm R}$ is not induced by $z$, as
expected for a sample with almost no flux limits in either bands.
The small dispersion
in the L$_{\rm R}$ vs. L$_{\rm X}$ correlation is remarkable, since the X-ray emission is 
known to be significantly variable in AGN (e.g. Mushotzky et al. 1993), and the radio observations, which also show some variability (Barvainis et al. 2005), were taken about 10 years
before the X-ray observations (early 80s vs. early 90s), suggesting that the intrinsic scatter
in the R-X relation is smaller than measured here. 

The tight relation between the radio and X-ray emission, despite
the disparity by a factor of $>10^7$ in energy, may provide
a hint for the origin of the radio emission in RQQ, as further discussed below.

\subsection{The Palomar-Ho et al. sample of low-luminosity AGN}

How far down in luminosity does the R-X relation extend?  Unfortunately, there is no 
comparable optically selected sample of low luminosity type 1 AGN with nearly complete X-ray, UV,
and radio detections, as the PG sample used above. However, the Palomar optical spectroscopic survey 
of nearby galaxies (Ho et al. 1995, hereafter PH sample) comes close. The sample includes 486 galaxies which were carefully classified by Ho et al. (1997), who found that 52 are Seyfert galaxies, of which 13 are of type 1.0 to type 1.5, i.e. show prominent broad Balmer lines,
and are therefore unobscured. 
One of these, NGC~1275 (3C~084) is a well known radio loud AGN, and we therefore exclude it
from the sample of low luminosity unobscured RQ AGN. Systematic analysis of Radio (Ho \& Ulvestad 2001) and X-ray (Panessa et al. 2006) observations is available for all 12 objects. There is no similar study of the UV spectra of these objects, but the spectra of 8 of the objects can be found in various studies
(Evans \& Koratkar, 2004; Maoz et al., 1998; Crenshaw et al., 1999; 2001; Kaspi et al. 2004) which often show some \c\ absorption. Given the small size of this sample, and the non complete UV coverage, we do not exclude objects based on the presence of UV absorption. One should keep in mind, however,
that some absorption may affect the $L_{\rm X}$ measurements 
(though heavily absorbed ``Compton thick" objects were excluded by Panessa et al. 2006). 
Also, X-ray variability, which tends to be more prominent in lower luminosity objects, will
contribute towards increasing the scatter in correlation studies for this population.

Figure 2b shows a ``zoom-out" of the R-X relation to include the 12 lower luminosity 
PH AGN. The values of $L_{\rm R}$ from Ho \& Ulvestad (2001) were corrected for the revised distances
in Panessa et al. (2006), excluding NGC~4395, where the more recent Thim et al. (2004) value of
4.3~Mpc is used. We also corrected $L_{\rm R}$ in Ho \& Ulvestad (2001) in objects where mas scale observations indicate that the VLA core is resolved. Specifically, we used
a flux of 8~mJy instead of 23.8~mJy in NGC~3227 (Mundell et al. 1995), 10~mJy instead of 77.6~mJy
in NGC~4151 (Ulvestad et al. 1998), and 21~mJy instead of 37.7~mJy in NGC~4579 (Hummel et al. 1987).
The 2-10~keV luminosities in Panessa et al. (2006) were converted to the 0.2-20~keV luminosities,
i.e. L$_{\rm X}$, by multiplying L$_{\rm 2-10~keV}$ by 2.86, which applies for a uniform energy
slope of $-1$. The typical spectral slopes are somewhat steeper below $\sim 1$~keV, and flatter above
$\sim 1$~keV, and the resulting bolometric correction error is not a significant source 
of scatter. 
Since NGC~4395 shows very large and rapid X-ray variability, we have used the time-averaged absorption-corrected L$_{\rm 2-10~keV}=8.8\times 10^{39}$~erg~s$^{-1}$, measured by
Moran et al. (2005, corrected for a distance of 4.3~Mpc). Clearly, it would be useful
to carry out a more complete study of the radio and X-ray emission properties of the PH sample.

The 12 PH AGN extend the R-X relation from the range of log L$_{\rm X}=43-45$ probed by
the PG sample, to the range of log L$_{\rm X}=40-43$. The solid line in Fig.2b is not a 
fit, but just a line marking L$_{\rm R}$/L$_{\rm X}=10^{-5}$,
a typical value for the PG quasars. The PH AGN sample together with the PG sample, suggest 
a tight linear R-X relation from the lowest luminosity type 1 AGN, NGC~4395 at
log~L$_{\rm X}\simeq 40$ (left most AGN in Fig.2b), up to luminous quasars at
log~L$_{\rm X}\simeq 45$, with a slope of unity (consistent with eq.1). 
This trend can only be taken as suggestive, as the PH sample is
rather fragmentary, and a much larger complete sample with high quality data is required.

\subsection{The Most Luminous RQ AGN}

There is yet no study of the X-ray and radio emission properties of a complete sample of 
the most luminous RQ AGN. However, some indications can be obtained by combining the results
of independent studies of the X-ray emission and of the radio emission in the most luminous
AGN (Steffen et al. 2006). More specifically, Just et al. (2007) find that $L_{\rm 2keV}\propto L_{\rm 2500\AA}^{0.64-0.81}$,
using a large compilation of AGN extending to $\nu L_{\nu}$(2500\AA)=$4\times 10^{47}$~erg~s$^{-1}$, 
about a factor of 100 more luminous than the RQ PG quasars. 
{The slope range of 0.64--0.81 comes from treating $L_{\rm 2keV}$ either as the dependent or the independent variable.
Thus, the X-ray to UV luminosity ratio
decreases with increasing luminosity. A similar trend of a decreasing radio to UV luminosity 
ratio was recently found by White et al. (2007) in stacking analysis of undetected luminous
quasars from the FIRST radio survey. They found 
$L_{\rm R}\propto L_{\rm 2500\AA}^{0.85}$ (with no error quoted for the slope) for AGN with a luminosity extending to
$\nu L_{\nu}$(2500\AA)=$6.5\times 10^{47}$~erg~s$^{-1}$ (corresponding to $M_{2500\AA}=-30.25$).
Thus, both the radio and X-ray emissions become relatively weaker with increasing UV luminosity.
Combining the two correlations above gives $L_{\rm R}\propto L_{\rm X}^{1.05-1.33}$,
i.e. marginally consistent with a slope of unity suggested for lower luminosity AGN.
Deep radio observations
of a complete sample of the most luminous AGN are required to probe directly the extension of the R-X relation to the highest luminosities.
All these studies of X-ray and radio luminosity relations find no redshift dependence, suggesting perhaps they are due to fundamental physics.

\subsection{Ultraluminous X-ray Sources}

At yet lower luminosities, possible counterparts to accreting massive black holes
are the Ultraluminous X-Ray Sources (ULXs). These are point like, off-nuclear sources, 
detected in nearby galaxies, with isotropic luminosities of up to log~$L_{\rm X}\simeq 41$. 
The high luminosity can be interpreted either as isotropic emission from intermediate-mass
black holes ($M_{\rm BH}\sim 10^2-10^3~M_{\odot}$) accreting at $L 
\sim L_{\rm Eddington}$, or as strongly beamed jet emission from stellar mass black holes,
presumably within a massive X-ray binary (e.g. Mushotzky 2006 for a recent review).

In Fig.2b we plot four ULXs that have 
candidate radio counterparts. Little information is available about the radio emission properties  of ULXs. Following a literature search we found four objects with relatively 
secure radio identifications. For simplicity, in all cases we extrapolated the unabsorbed 
X-ray luminosities reported in the
literature to the full 0.2-20~keV band assuming an energy spectral  
slope of $-1$ (also used for the radio).
In most cases, the unabsorbed X-ray luminosities are obtained from  
fitted models with column
densities above Galactic, which could introduce further uncertainty.
However, this uncertainty is likely small compared with the large dynamic range
covered in Fig. 2b.

For the ULX in NGC\,5408 (Kaaret et al. 2003) we infer
$L_{\rm X}=1.5\times 10^{40}$~erg~s$^{-1}$ and $L_{\rm R}=3.5\times  
10^{34}$~erg~s$^{-1}$.
For $L{\rm _X}$ of the ULX in Holmberg~II we take the average of the  
low and high states
to obtain $L_{\rm X}=1.7\times 10^{40}$~erg~s$^{-1}$ (Dewangan et al.  
2004) and
$L_{\rm R}=4.7\times 10^{34}$~erg~s$^{-1}$ (Miller et al. 2005).
The third and most luminous ULX, X-1 in M\,82 (Kaaret et al. 2006),  
was observed
simultaneously with {\it Chandra} and the VLA.
The luminosities obtained are
$L_{\rm X}=3\times 10^{40}$~erg~s$^{-1}$ (extrapolated from  
0.3--7.0~keV) and
$L_{\rm R}=6.7\times 10^{34}$~erg~s$^{-1}$ (on 2005 Feb. 5).
We note that the radio identification of this ULX was questioned by  
K{\"o}rding et al. (2005).
Another likely association was found for ULX~2 in NGC\,7424 (Soria et al. 2006), where
$L_{\rm X}=1.06\times 10^{40}$~erg~s$^{-1}$ (using the power-law model),
and $L_{\rm R}=1.05\times 10^{35}$~erg~s$^{-1}$. The radio
source is offset by $\simeq 1.4$~arcsec, or 80~pc, from the X-ray  
position, but given the
lack of other nearby point sources, the probability of a chance  
coincidence is low (Soria et al. 2006).

Interestingly, the four ULXs are located  
close to the
$L_{\rm R}$/$L_{\rm X}=10^{-5}$ line (Fig.2b).
Since the radio and X-ray emission of the PG and PH AGN are most  
likely unbeamed, this may
imply that the ULX radio and X-ray emission is unbeamed as well, as  
indicated by various
other arguments (Mushotzky 2006). Furthermore, this may indicate that  
similar physical
processes generate the radio and X-ray emission in ULXs and in AGN.
Clearly, this is a very small sample of ULXs and more radio  
identifications and reliable flux measurements will be
required to more systematically constrain the $L_R/L_X$ behavior of  
these intriguing sources.
In particular, it should be interesting to see if there are also ULXs  
significantly below the G{\"u}del-Benz relation
(i.e., with hot disks akin to GBHs, see section 2.6) or above it (i.e.,  
similar to radio-loud AGN).

\subsection{Coronally Active Stars}

A strong correlation is known to exist between the quiescent radio and X-ray emission in 
coronally active cool stars (G{\"u}del \& Benz 1993), 
where L$_{\rm R}$/L$_{\rm X}\simeq 10^{-5}$, also known as the G{\"u}del-Benz relation. 
In Figure 2c we further increase the luminosity range of the R-X plot to
include coronally active stars.
We plot a compilation of radio and X-ray luminosities of various types of coronally active stars 
from Benz \& G{\"u}del (1994). Quite remarkably, {\em both coronally active
stars and active galaxies appear to follow a similar R-X relation}, despite the 
$\sim 15$ orders of magnitude in luminosity between these two types of active objects.

The X-ray emission in coronally active stars is due to free-free and line 
emission from hot thermal plasma (T$\sim 10^7$K), while the radio is synchrotron emission
from non-thermal electrons embedded in a magnetic field. The stellar R-X relation indicates 
that the coronae are magnetically heated, as also indicated 
by correlated X-ray and radio variability. 
Note that the data points from the Sun are for different kinds of solar {\it flares} that may deviate from the general stellar coronae linear relation (Benz \& G{\"u}del 1994).
The R-X relation can be understood 
as follows. A magnetic reconnection 
event in the corona accelerates electrons, generating a beam of electrons 
which produces a spike in the synchrotron radio emission. 
The beam dumps most of its energy by collisions 
in the cooler chromospheric gas, heating it to coronal temperatures, and thus increasing the
total coronal X-ray emission. Strong support for this mechanism is provided
 by the Neupert effect (Neupert 1968, originally observed on the Sun), where a radio
flare is followed by an X-ray flare, where the two light curves are related through
\begin{equation} 
L_{\rm R}\propto dL_{\rm X}/dt 
\end{equation} 
(G{\"u}del 2002, and references therein).
Such a relation is expected when the X-ray cooling time is significantly longer than the radio cooling time, so the X-ray emitting gas serves as a bolometer for the time integrated heating, while the radio emission provides a measure for the instantenous heating rate.

The fact that RQ AGN and coronally active stars follow a similar R-X relation may indicate
similar physical processes in both objects, as further discussed in section 3.

\begin{figure*}
\includegraphics[width=165mm,angle=0]{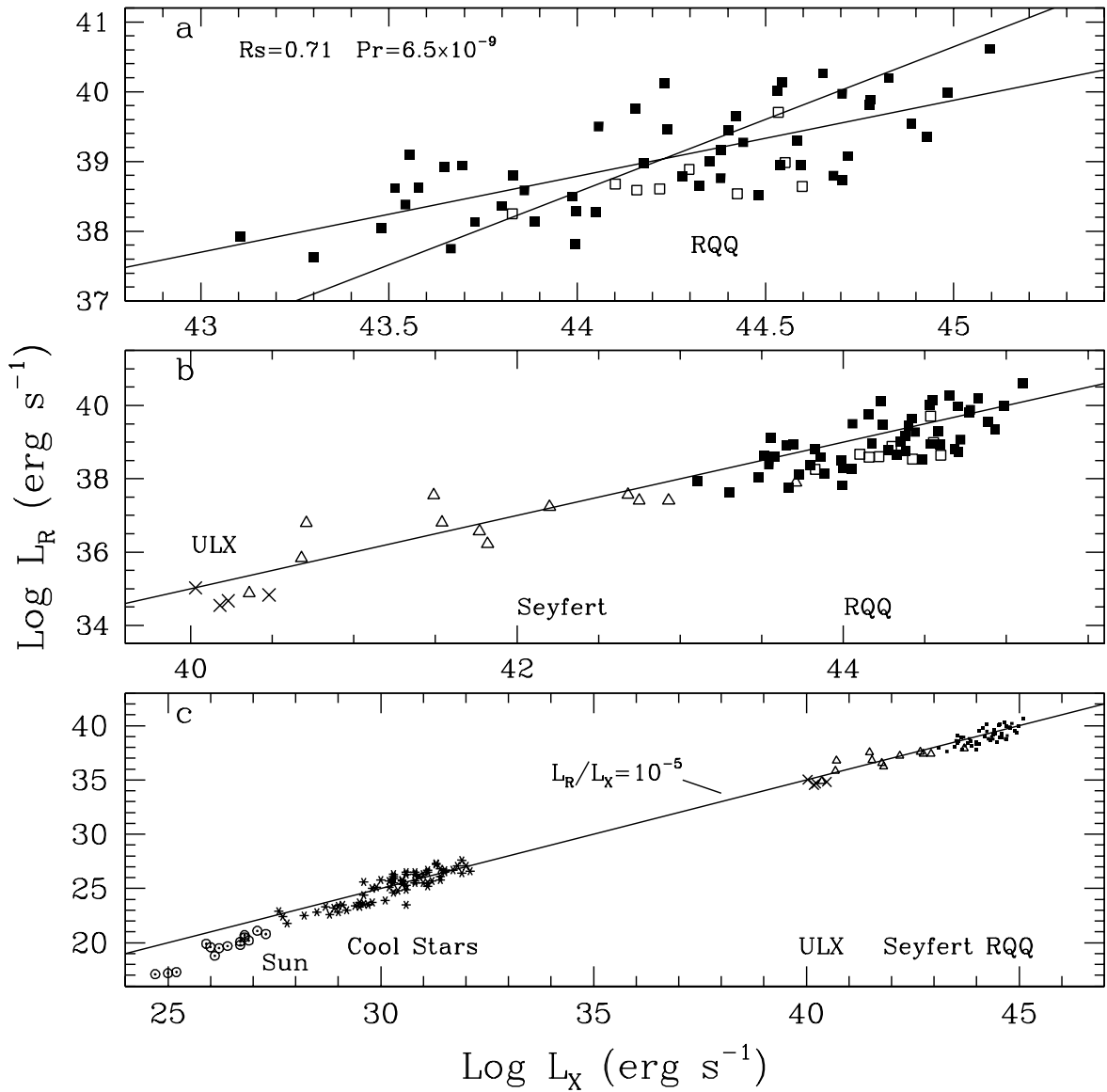}
\caption{Panel a, same as in Fig.1, including only the RQ AGN with no UV absorption.
The Spearman rank order correlation strength and its significance is indicated in the
upper left corner. The two lines mark least square fits minimizing the deviations in
either $L_{\rm R}$ or $L_{\rm X}$. Panel b, zoom out which includes the 12 low luminosity
RQ type 1 AGN from the Palomar sample, and four ULXs with radio and X-ray detections. 
Most objects are within a factor
of few from the solid line which marks $L_{\rm R}/L_{\rm X}=10^{-5}$.
Panel c, further zoom out to include coronally active stars, and individual solar 
flares. Coronally active stars follow the G{\"u}del-Benz relation, i.e. 
$L_{\rm R}/L_{\rm X}=10^{-5}$, which most likely originates from coronal radio and X-ray
emission. This 
raises the possibility that both $L_{\rm R}$ and $L_{\rm X}$ in RQ AGN also originate in
coronal emission.}
\end{figure*}

\begin{figure*}
\includegraphics[width=165mm,angle=0]{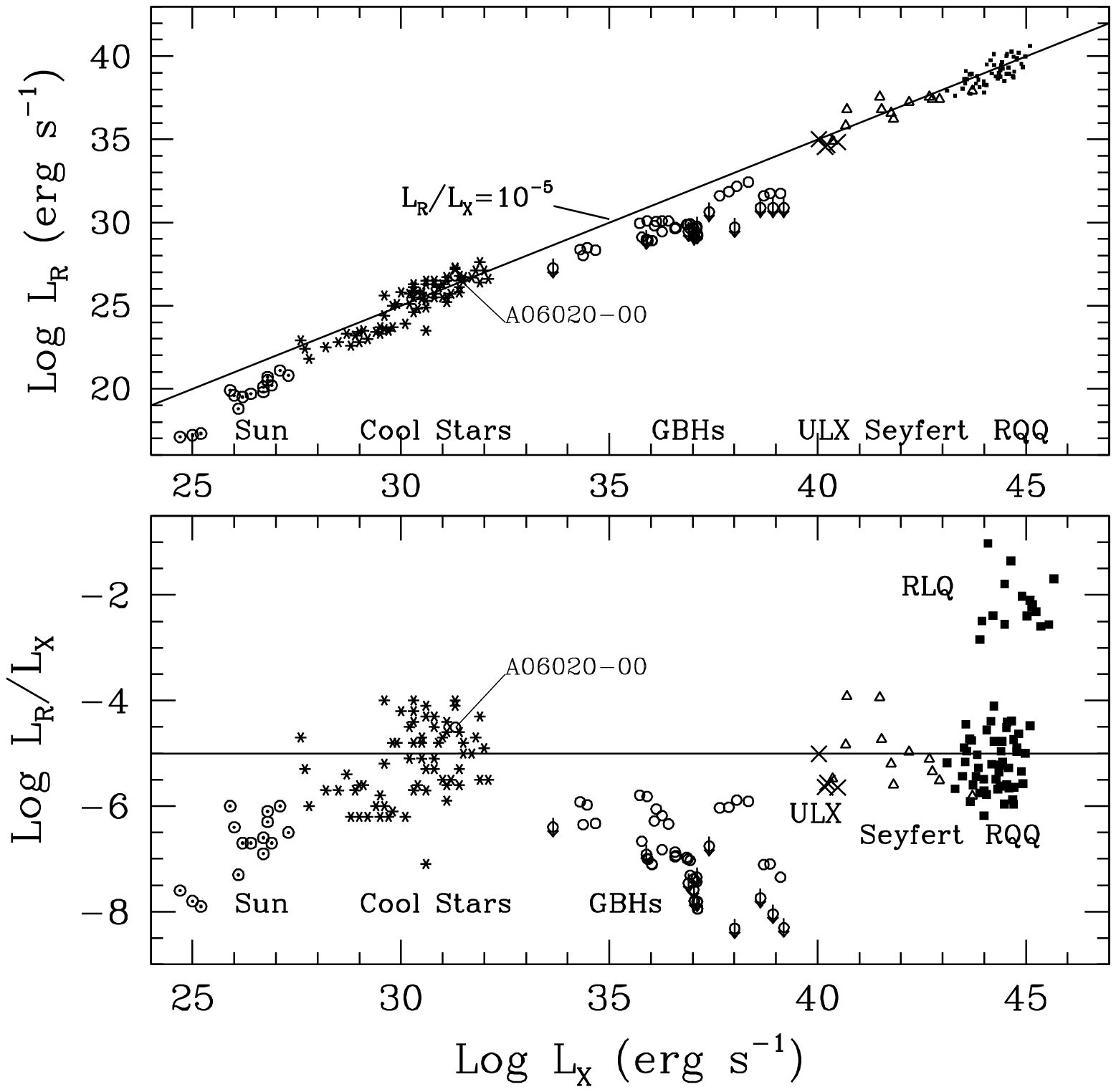}
\caption{Upper panel, the position of GBHs in the $L_{\rm R}$ vs. $L_{\rm X}$ plane. The data include
eight GBHs in the low hard state (including repeat observations), taken from the compilation of
Merloni et al. (2003). All GBHs are located downwards from the $L_{\rm R}/L_{\rm X}=10^{-5}$
relation by a factor of $10-1000$. GBHs are thus more radio quiet than radio quiet AGN. 
An additional GBH, A0620-00, detected at the quiescent state by Gallo et al. (2006) is also 
included. At Log $L_X=31.31$ and Log $L_R=26.80$ it overlaps with the coronally active stars.
Lower panel, the distribution of $L_{\rm R}/L_{\rm X}$ values, including RLQ. Note the similar
distribution of values for RQQ and coronally active stars.}
\end{figure*}

\subsection{Galactic Black Holes}

The R-X relation was also extensively explored in massive Galactic X-ray binary systems
(GBHs, e.g. Merloni et al. 2003; Falcke et al. 2004). Do
GBHs also follow the G{\"u}del-Benz relation? In Figure 3 we show the R-X relation for eight GBHs, 
in the low hard state, from the compilation of Merloni et al. (2003, including multiple observations of some of the 
objects). The GBHs are clearly displaced downwards from the G{\"u}del-Benz relation, having
L$_{\rm R}$/L$_{\rm X}\simeq 10^{-7}$-$10^{-6}$, or less. In other words, 
GBHs are more radio quiet than RQ AGN.
The radio and X-ray emission of GBHs in the low hard state is commonly interpreted as 
arising from jet emission,
although spatially resolved jets are generally not observed. Spatially resolved jets
in RL AGN often display L$_{\rm R}$/L$_{\rm X}\ga 10^{-2}$ (e.g. Miller et al. 2006b;
Sahayanathan \& Misra 2005; Hardcastle et al. 2007), as also observed for the spatially
integrated radio and
X-ray emission in these objects (Panessa et al. 2007; Capetti \& Balmaverde 2007).
Thus, the jets in GBHs should be different from those in radio loud AGN, possibly due to a strong 
$M_{\rm BH}$ dependence of the radio/X-ray emission ratio of jets (e.g. Merloni et al. 2003; 
Falcke et al. 2004). Another possible model invokes advection-dominated accretion with a truncated
inner disk (recently reviewed by Narayan \& McClintock 2008).
Alternatively, if the X-ray and radio emission of GBHs in the low hard state
does originate
in an accretion disk corona, the properties of GBH coronae should be somewhat different from
the coronae of AGN and of coronally active stars.

We note in passing, that the least luminous GBH observed, A$0620-00$, detected at the quiescent state by Gallo et al. (2006), with Log L$_{\rm X}$=31.31 and Log L$_{\rm R}$=26.80, overlaps in Fig.3
with the coronally active stars. Gallo et al. argued that the radio and X-rays cannot be coronal emission from the companion star, based on its spectral type. The overlap of this object with coronally active stars may
provide an important clue to understanding the origin of the radio and X-ray emission in
GBHs, as further discussed below (section 3.3.1).

\section{Some Implications}

\subsection{The Radio Emission Isotropy}

The small scatter in the R-X relation for RQ AGN, described above, implies that the angular
distribution of the emission in both bands is not significantly different. The strength of 
the X-ray emission in AGN is correlated with the broad line emission strength 
(e.g. Kriss et al. 1980). Since the broad line emission cannot be relativistically beamed, 
the X-ray emission is generally not relativistically beamed as well. Indeed, the X-ray emission is
commonly thought to originate from relatively static hot coronal gas above the accretion disk 
(see below). Thus, we conclude that the radio emission in RQ AGN is also generally not relativistically 
beamed. If the radio emission of RQ AGN occurs in jet emission, then the jet cannot be highly relativistic, as seen in the radio for some RL AGN. 

Blundell et al. (2003) suggested a detection of superluminal motion in a RQQ
based on high brightness temperature. However, 
extended structures in VLBA observations of other RQQs were not
confirmed by Ulvestad et al. (2005), indicating that 
there is yet no direct evidence for superluminal motion in RQQs.

\subsection{The analogy between AGN and coronally active stars}

The power-law X-ray emission of AGN is distinctly non-thermal, i.e. it does not originate 
from free-free emission of photoionized or collisionally ionized plasma. The energetically 
least demanding non-thermal mechanism is
Comptonization of the optical-UV disk continuum by hot thermal ($T\ga 10^9$~K), 
or non-thermal 
electrons residing in a magnetically confined gas above the accretion disk 
(e.g. Shapiro et al. 1976; Sunyaev \& Titarchuk 1980). The existence of magnetically confined
and heated hot  gas above the accretion disk is plausible given the differential 
rotation in Keplerian accretion disks, and the likely presence of turbulence 
(required to generate macroscopic viscosity). These conditions are similar to 
those in coronally active stars, which are 
rapidly rotating and have a turbulent surface layer (reviewed by G{\"u}del 2004). 
Furthermore, magnetic fields within the accretion 
disk are invoked as a plausible viscosity mechanism
(Eardley \& Lightman 1975; Balbus \& Hawley 1998), and the rise of buoyant flux tubes
from the disk can serve as a natural confining mechanism of the hot coronal gas,
while magnetic reconnection can serve as a natural coronal heating mechanism
(Galeev et al. 1979; Blackman \& Field 2000), again  in analogy to observations of 
Coronally active stars. 

Observational implications of the X-ray emitting corona model have been worked out
in relation to the expected X-ray continuum (e.g. Haardt \& Maraschi 1993; Dove et al. 1997;
Nied{\'z}wiecki 2005), 
X-ray spectral features (e.g. Ross \& Fabian 1993; Matt et al. 1997; 
Nayakshin et al. 2000; Ballantyne et al. 2001) and
X-ray variability (e.g.  Poutanen \& Fabian 1999; Merloni \& Fabian 2001b; Czerny et al. 2004). 
However, little attention was given to the emission at
longer wavelengths. Notable exceptions are Field \& Rogers (1993), di Matteo et al. (1997), 
and Merloni et al. (2000) who addressed
the inevitable cyclotron or synchrotron emission from the magnetically confined 
thermal and non-thermal electrons, though none of these papers attempted to ascribe 
the observed radio emission of RQ AGN to coronal activity.

The fact that AGN follow the G{\"u}del-Benz relation, seen in objects where
both the radio and X-ray emission are of coronal origin, together with the fact 
that the X-ray emission in AGN is likely of coronal origin, leads naturally to 
the suggestion that the radio emission 
in RQ AGN is also coronal in origin.

Typical luminous AGN are approximately $13-14$ orders of magnitude more luminous than 
stellar coronae (Fig.~2),  {\em why should similar physics apply to stellar and AGN coronae?} 
Although luminous AGN are vastly more luminous than stars, their peak emission generally
occurs in the near UV (e.g. Zheng et al. 1997), indicating maximal 
surface disk temperatures of $T_s\sim 5\times 10^4$~K, within 
an order of magnitude of the photospheric temperatures of coronally active stars.
The dimensions of the X-ray emitting region in luminous AGN
is of the order of a light day, i.e. 
$\sim 10^{15}$~cm, corresponding to a volume of $\sim 10^{45}$~cm$^3$.
Stellar coronal volumes, on the other hand, are of the order of 
$0.01R_*^3 - 0.1R_*^3 \sim 10^{31} - 10^{32}$~cm$^3$, based on loop half-length measurements  
($l \sim R_*$) and cross-section radius estimates ($r \sim 0.1l$).
Thus, the coronal volume in AGN is likely to be
$\sim 13-14$ orders of magnitude larger than in stars.
If the coronal luminosity is proportional to the coronal volume $\times B^2$,
then the value of $B^2$ may be comparable in stellar and in AGN coronae. The 
potential similarity
of $T_s$ and $B^2$ in stellar and AGN coronae may then imply similar local physical 
processes, despite the vast difference in scales.

\subsection{Is the X-ray and radio emission of GBHs also of coronal origin?}

\subsubsection{The dependence of L$_{\rm R}$/L$_{\rm X}$ on $T_s$}

The radio emission of GBHs lies a factor of 10-100 below the G{\"u}del-Benz relation.
{\em Can the radio and X-ray emission of GBHs also be explained as coronal emission?}
Merloni \& Fabian (2002) suggested that the level of coronal activity in GBHs is related 
to $T_s$, such that the coronal activity diminishes with increasing $T_s$, as indicated by
the disappearance of the X-ray power law component and the radio emission in the high 
soft state of GBHs. One may extend this suggestion, and speculate that the overall low 
$L_{\rm R}/L_{\rm X}$ in GBHs is related to their overall high accretion disk 
$T_s$ compared to AGN.
Specifically, the effective accretion disk surface temperature at a given dimensionless
$r=R/R_g$, where $R_g\equiv GM_{\rm BH}/c^2$,
is 
\begin{equation} 
T_{\rm eff}\propto (\frac{\dot{m}}{r^3m})^{1/4}, 
\end{equation} 
where $\dot{m}\equiv L/L_{\rm Eddington}$,
and $m=M_{\rm BH}/M_{\odot}$ (e.g. Shakura \& Sunyaev 1973).
In luminous GBHs $\dot{m}/m\sim 10^{-2}$, while
in luminous AGN $\dot{m}/m\sim 10^{-10}$, implying that $T_s$ in GBHs
is $\sim 100$ times higher than in AGN (assuming $T_s\simeq T_{\rm eff}$), 
as indeed observed (peak emission at keV in GBHs
versus tens of eV in AGN). Thus, one may speculate that the 
$L_{\rm R}/L_{\rm X}$ in the corona scale roughly as  $T_s^{-1}$, 
leading to $\sim 100$ times lower 
$L_{\rm R}/L_{\rm X}$ in GBHs
compared to AGN. A potentially key object in that respect is A$0620-00$,
the least luminous GBH detected (section 2.6), which is $\sim 10^7$ times less luminous than typical luminous
GBHs. In this object $\dot{m}/m\sim 10^{-9}$, and thus if it harbors an optically
thick accretion disk, as indicated by observations of a broad fluorescence
Iron $K\alpha$ line in other
low state GBHs (Miller et al. 2006a, Rykoff et al. 2007; but see counter
arguments for a truncated inner disk in a review by Done et al. 2007, section 4.3 there), 
then its $T_s$ will be similar to that of AGN. Indeed, this GBH
lies on the G{\"u}del-Benz relation (Fig.3), consistent with the suggestion that 
$L_{\rm R}/L_{\rm X}$ is set by $T_s$.

We note in this context that if ULXs are powered by intermediate mass BHs 
($m\sim 10^3-10^4$) they may also be expected to be intermediate in their 
$L_{\rm R}/L_{\rm X}$ between luminous AGN and GBHs. The scant available data (Fig.2) indicates
they may lie somewhat below the G{\"u}del-Benz relation, but the scatter is too large to reach
any solid conclusion. 

Interestingly, Merloni et al. (2003) found a "fundamental plane" relation
connecting GBHs and AGN through
$L_{\rm R}\propto L_{\rm X}^{0.6}m^{0.78}$ (see also Falcke et al. 2004). 
Their sample includes 5 RL AGN, and 23 Seyfert 2 galaxies. Objects of this type are not
present in our sample. When these objects are excluded one gets a slightly
modified relation
\begin{equation}
L_{\rm R}\propto L_{\rm X}^{0.54\pm 0.14}m^{0.78\pm 0.13}, 
\end{equation} 
(A. Merloni, private communication). This relation can be recast in the form 
\begin{equation} 
L_{\rm R}/L_{\rm X}\propto L_{\rm X}^{-0.46\pm 0.14}m^{0.78\pm 0.13}. 
\end{equation}
Using the Just et al. (2007) relation $L_{\rm X}\propto L_{\rm UV}^{0.64-0.81}$,
or roughly $L_{\rm X}\propto L_{\rm UV}^{0.725\pm 0.085}$,
and assuming $L_{\rm bol}\propto L_{\rm UV}$, which appears to apply
in both high and low luminosity AGN (Maoz 2007), we get
\begin{equation} 
L_{\rm R}/L_{\rm X}\propto L_{\rm bol}^{-0.33\pm 0.11}m^{0.78\pm 0.13},
\end{equation}
or
\begin{equation}
L_{\rm R}/L_{\rm X}\propto (\dot{m}/m)^{-0.33\pm 0.11}m^{0.12\pm 0.26}. 
\end{equation}
where we use $\dot{m}\propto L_{\rm bol}/m$.
For a Shakura \& Sunyaev accretion disk, 
this can be recast in the form
\begin{equation} 
L_{\rm R}/L_{\rm X}\propto T_{\rm eff}^{-1.32\pm 0.44}m^{0.12\pm 0.26}, 
\end{equation} 
i.e. consistent with the suggestion that the radio and X-ray emissions
are both coronal in origin and that 
$L_{\rm R}/L_{\rm X}\propto T_{\rm eff}^{-1}$. Thus, $T_{\rm eff}$ 
may be the physical parameter 
which drives the "fundamental plane" 
relation noted by Merloni et al. (2003) and Falcke et al. (2004).

We do not see a significant correlation between $L_{\rm R}/L_{\rm X}$ and
$T_{\rm eff}$, or equivalently we do not see the Merloni et al. relation for the PG 
sample only. This may be due to the small range in values 
of $T_{\rm eff}$ in this sample (a factor of $\sim 10$) compared to the large scatter 
in $L_{\rm R}/L_{\rm X}$ values (a factor of $\sim 100$, Fig.3), which makes it difficult
to see such a trend, if it exists.
Interestingly, it is hard to discern X-ray (or radio) luminosity dependence on spectral type in stellar coronae as well, perhaps due to the small range of cool-star photospheric temperatures (but see the recent result of Scelsi et~al. 2007).
The scatter in $L_{\rm R}/L_{\rm X}$ values must be partly due to the non-simultaneous X-ray and radio observations, and may be partly intrinsic as well.
The "fundamental plane" relation is not detectable for the GBH-only
population either (Fig.3), most likely for the same reasons.
Interestingly, individual GBHs often show a non-linear relation 
$L_{\rm X}\propto L_{\rm R}^{0.7}$ in a given object (e.g. Corbel et 
al. 2000; Gallo et al. 2003, 2006),
consistent with the ``fundamental plane" relation at a fixed $m$ (eq.5). It would thus be
interesting to explore if this non-linear relation is seen in individual AGN as well.

\subsubsection{Correlated variability}

Malzac et al.(2003) found an intriguing pattern of variability in the GBH
XTE J1118+480, where 
\begin{equation} 
L_{\rm opt}\propto -dL_{\rm X}/dt\ .
\end{equation}
The optical luminosity is interpreted as synchrotron emission from a jet,
and the X-ray as coronal emission. This unusual variability pattern is
reminiscent of the Neupert effect (section 2.5), seen in coronally 
active stars, where $L_{\rm opt}$ is replaced by $L_{\rm R}$, and
the sign in eq. (3) is positive, rather than negative.
Malzac et al.(2004) suggested a model to explain this pattern, where both the 
jet and the corona are powered by a common magnetic energy reservoir. 
The Neupert effect is interpreted as the signature of a coronal reconnection
event. The``Malzac effect" may thus be driven by similar processes, and it
may therefore be a pure coronal effect with no need to invoke a jet component.
The relativistic electrons may be part of the corona, which may be the agent through
which they are heated, rather than reside in a separate jet component. However, 
a significant difference between the two effects is in the sign of the derivative,
and the fact that both flares and dips are observed in GBHs, but only flares
are seen in stellar coronae. These differences remain to be understood.

\subsection{The jet interpretation}
The currently favored interpretation for the R-X relation in GBHs and
in AGN is that it results from
a disk/jet coupling (e.g. Merloni et al. 2003; Falcke et al. 2004) or more
specifically from a corona/jet coupling,
as the corona feeds the base of the jet (e.g. Merloni \& Fabian 2002; 
Markoff et al. 2005).  
If RQ AGN are powered by weak jets, then it is also not clear why RQ AGN are often
unresolved even on mas scale, in clear contrast to RL AGN which are generally well
resolved already on much larger scales. In addition, the low power jet interpretation does 
not provide a natural explanation as to why AGN appear to follow the G{\"u}del-Benz
relation. The above does not rule out the low-power jet interpretation for RQ AGN, 
but the coronal
interpretation appears as a viable alternative.

\subsection{The size of the radio emitting region}

\subsubsection{The synchrotron emission mechanism}

Below we derive a lower limit on the size of the radio emitting region in
RQ AGN, based on the maximum flux per unit area emitted by a synchrotron source. 
The observed flux density from an isotropically emitting source is
\begin{equation} 
f_{\nu}=I_{\nu}\Omega ,
\end{equation}
where $I_{\nu}$ is the intensity (independent of angles), and
$\Omega$ is the angular size of the source. For a source with a projected
shape of a circle with a radius $R$, $\Omega=\pi R^2/d^2$, where $d$ is the angular size 
distance. In a homogeneous source
\begin{equation} 
I_{\nu}=S_{\nu}(1-e^{-\tau_{\nu}}) ,
\end{equation}
where $S_{\nu}\equiv P_{\nu}/4\pi\alpha_{\nu}$ is the source function,
$P_{\nu}$ the power emitted per unit volume per unit frequency
and $\alpha_{\nu}$ is the absorption coefficient. The minimum emitting
area required to produce an observed $f_{\nu}$ is obtained when $I_{\nu}$ is 
maximal, i.e. $I_{\nu}=S_{\nu}$ which is obtained
when the source is optically thick. Below we briefly review the
derivation of $S_{\nu}$ 
based on the theory of synchrotron emission formulated by Ginzburg
\& Syrovatskii (1969) and later presented by Rybicki \& Lightman 
(1979, hereafter RL).

We assume a homogeneous synchrotron source, with a uniform magnetic field $B$,
and a uniform distribution of relativistic electrons, having a power-law 
energy distribution of the form
\begin{equation} 
dn/d\gamma=C_{\gamma}\gamma^{-p}, \ \ \ \gamma_0< \gamma< \gamma_1 .
\end{equation}
The electrons radiate synchrotron emission with a power
per unit volume per unit frequency of (equation 6.36 in RL),
\begin{equation} 
P_{\nu} = \frac{\sqrt{3}q^3}{mc^2(p+1)}
\left( \frac{3q}{2\pi mc} \right)^{\frac{p-1}{2}}C_{\gamma} B_\perp^{\frac{p+1}{2}} 
\nu^{-\frac{p-1}{2}} \Gamma_1 \Gamma_2 ,
\end{equation}
where $B_\perp$ is the component of $B$ perpendicular to the direction
of motion of the electrons, and
\[\Gamma_1=\Gamma \left( \frac{3p+19}{12} \right) ,\ \ \
\Gamma_2=\Gamma \left( \frac{3p-1}{12} \right) \]
The absorption coefficient is given by
\begin{equation} 
\alpha _\nu = \frac{\sqrt{3}q^3}{8\pi c^2m^2}
\left( \frac{3q}{2\pi mc} \right)^{\frac{p}{2}}C_{\gamma} 
B_\perp^{\frac{p+2}{2}} \nu^{-\frac{p+4}{2}}
\Gamma_3 \Gamma_4 ,
\end{equation}
as derived from equation 6.53 in RL, by replacing $C$ from equation 6.20a there
(G. Rybicki, private communication), with $C_{\gamma}$. Here
\[\Gamma_3=\Gamma \left( \frac{3p+22}{12} \right) ,\ \ \
\Gamma_4=\Gamma \left( \frac{3p+2}{12} \right) .\]
The synchrotron source function is therefore
\begin{equation} 
S_{\nu}=\frac{1}{p+1}\sqrt{\frac{8\pi m^3c\nu^5}{qB_\perp}}
\frac {\Gamma_1 \Gamma_2}
{\Gamma_3 \Gamma_4 } ,
\end{equation}
or
\begin{equation} 
S_{\nu}=\frac{6.29\times 10^{-31}}{p+1}B_\perp^{-1/2}\nu^{5/2} R_{\Gamma}~
{\rm erg~s}^{-1} {\rm cm}^{-1} {\rm Hz}^{-1} {\rm Ster}^{-1}
\end{equation}
where the ratio of the $\Gamma$ functions $R_{\Gamma}=1.37$, for $p=2$, and
1.197 for $p=2.5$. The luminosity density of the source is 
$L_{\nu}=f_{\nu}4\pi d^2$, which together with equation 11 gives the radius
of the synchrotron radio-sphere
\begin{equation} 
R^{\rm RS}=\frac{1}{2\pi}\sqrt{\frac{L_{\nu}}{S_{\nu}}}
\end{equation}
where we assume that the luminosity and angular size distances, $d$, are the same 
(a good approximation for the low-$z$ objects studied here). Thus the minimum size
of the area emitting $L_{\nu}$ in synchrotron emission from a homogeneous source
with a given $B_\perp$, and electrons with $p=2$ is
\begin{equation} 
R_{\rm pc}^{\rm RS}=0.54 L_{30}^{1/2}\nu_{\rm GHz}^{-5/4}B^{1/4}\ , 
\end{equation}
where  $R_{\rm pc}^{\rm RS}$ is the radius of the radio-sphere in pc,
$L_{30}=L_{\nu}/10^{30}$~erg~s$^{-1}$~Hz$^{-1}$, and 
$\nu_{\rm GHz}$ is the observed frequency in GHz. For
$p=2.5$ the prefactor is 0.63, instead of 0.54. 

In the special case where the magnetic energy density 
is in equipartition with the photon energy density,
\begin{equation} 
B_{\rm eq}^2/8\pi=L_{\rm bol}/4\pi R^2c ,
\end{equation}
where $L_{\rm bol}$ is the bolometric luminosity,
the magnetic field can be estimated by:
\begin{equation} 
B_{\rm eq}=0.27R_{\rm pc}^{-1}L_{46}^{1/2}~{\rm Gauss} ,
\end{equation}
where $L_{46}=L_{\rm bol}/10^{46}$. Assuming $B=B_{\rm eq}$ in equation 19
then gives
\begin{equation} 
R_{\rm pc}^{\rm RS}=0.47L_{30}^{0.4}L_{46}^{0.1}\nu_{\rm GHz}^{-1} . 
\end{equation}
Note that $B\propto R^{-1}$ as in eq.21 is also expected if there is a large scale
current flow along the disk axis (the field of a wire). A steeper dependence, $B\propto R^{-2}$,
is expected for flux freezing in an expanding magnetized outflow. 

\subsubsection{The implied range of sizes for the emitting region}

Does the observed radio spectral slope indicates that the emitting source
is optically thick? A homogeneous synchrotron source, 
with a semi-infinite slab geometry, produces in the 
optically thick limit a spectrum with
$f_{\nu}\propto \nu^{2.5}$ (eq.17), and in the optically thin limit
$f_{\nu}\propto \nu^{-\frac{p-1}{2}}$ (eq.14), or $f_{\nu}\propto \nu^{-0.5}$
for a likely power-law slope $p=2$ of the radiating
electrons. The observed spectral slopes in RQQ at $\nu\sim 1-10$~GHz are in the 
range $-1<\alpha<1$ (e.g. Barvainis et al. 1996, 2005; Ulvestad et al. 2005),  
indicating that the radio source is either optically thin or 
is marginally optically thick. 
Thus 
$R_{\rm pc}^{\rm RS}\ga 0.1$, or about 100 light days, in luminous ($L_{30}\sim 1$, 
$L_{46}\sim 1$) AGN, 
with a steep spectral slope
at $\nu_{\rm GHz}=5$ (using eq.22). 
This radius is $\sim 100$ times larger than the likely size of the smallest
X-ray 
emitting region in such objects ($\sim 1$ light day based on variability). 
Similarly, in the lowest
luminosity type 1 AGN ($L_{30}\sim 10^{-5}$, $L_{46}\sim 10^{-5}$) 
$R_{\rm pc}^{\rm RS}\ga 3\times 10^{-4}$, or a light crossing time of $\ga 30$~ks, which
is also $\sim 100$ times larger than the smallest X-ray emitting region, as indicated by
the most rapid variability seen in e.g. NGC~4395 (Iwasawa et al. 2000;
Moran et al. 2005). The similar ratio of $100$ for the 5~GHz to X-ray emitting sizes
in high and low luminosity AGN reflects the fact that the fastest X-ray variability
appears to scale as $L^{1/2}$, similar to the scaling predicted for the 
radio (eq. 19).

The fact that the radio spectra never show a clear optically thick signature,
i.e. a spectral slope $\alpha=2.5$, despite the fact that this is a relatively 
robust prediction,
independent of the 
details of the electrons energy distribution, indicates that the
simple homogeneous source scenario is not valid. A plausible simple extension 
of the homogeneous source model is a 
spherical source with a radial gradient in both $B$ and in the number density 
of the relativistic electrons (e.g. de Bruyn 1976; Marscher 1977). 
In photospheric emission 
the flux at a given $\nu$ comes mostly from the region 
which is transiting from optically thick to optically thin at that $\nu$.
Using equation 19, we find that the transition frequency $ \nu_0$ is
\begin{equation} 
\nu_{0,{\rm GHz}} =0.61 L_{30}^{2/5}(R_{\rm pc}^{\rm RS})^{-4/5}B^{1/5}\ ,
\end{equation}
or for $B=B_{\rm eq}$
\begin{equation} 
\nu_{0,{\rm GHz}}=0.47L_{30}^{0.4}L_{46}^{0.1}(R_{\rm pc}^{\rm RS})^{-1}  .
\end{equation}
Thus we can relate each emission radius with a corresponding frequency
$\nu_0$. Since the synchrotron self-absorption coefficient
$\alpha_{\nu}\propto \nu^{-(p+4)/2}\sim \nu^{-3}$, higher frequencies originate
at smaller radii, and the emissivity integrated over the emitting volume can
produce a flat spectrum, as invoked in RL AGN (e.g. Blandford 
\& Konigl 1979 and citations thereafter). To probe synchrotron emission 
coming from the X-ray emitting volume, say from a region which is $\sim 100$ times smaller
than the $\sim 5$~GHz emitting region, requires a frequency 100 times higher,
i.e. $\sim 500$~GHz, or $\lambda \sim 0.6~mm$
(assuming $L_{\nu}\sim $ constant). 

\subsubsection{The shortest wavelengths dominated by synchrotron emission}
Is the emission at $\lambda \sim 1.2~mm$ in
RQ AGN dominated by synchrotron emission?
The FIR emission of RQ AGN is dominated by dust (e.g. Sanders et al. 1989), and
one needs to go to long enough wavelengths to make sure that dust contamination
of the synchrotron emission is insignificant. The observed mean spectral energy 
distribution (SED) of RQ AGN gives $L_{\rm IR}/L_{\rm R}\sim 10^6$, where
$L_{\rm IR}=\nu L_{\nu}(30\mu m)$. The coldest dust is found at $T\sim 30$~K
(e.g. Hass et al. 2000), and
the dust emission drops steeply at $\lambda >100~\mu m$ with a spectral slope 
for $L_\nu$ of $2+\kappa$, where $\kappa\simeq 2$ is the characteristic dust absorption opacity 
index (e.g. Draine \& Li 2007). The observed mean SED of RQ AGN indeed shows 
a sharp break with a $10^5$ drop in $\nu L_{\nu}$, from $100~\mu m$ to 1~mm 
(e.g. fig.7 in Polletta et al. 2000). Thus, dust contamination should be gone
at $\lambda\ga 1$~mm, and the mm wavelength range can provide a valuable
direct probe of the relativistic electron population within the  
X-ray emitting region.

\subsubsection{Constraints from the brightness temperature}

An independent constraint on the minimum possible size of a synchrotron emitting region can
be obtained using the source brightness temperature $T_b$, given by
\begin{equation} 
T_b=\frac{2c^2}{\pi k_{\rm B}}\frac{f_{\nu}}{\theta^2\nu^2}\ ,
\end{equation}
where $\theta=2R/d$ is the angular diameter of the source.
Recasting $f_{\nu}/\theta^2$ in terms of luminosity and 
emitting radius, we get

\begin{equation} 
T_b=8.7\times 10^9L_{30}(R_{\rm pc}^{\rm RS})^{-2}\nu_{\rm GHz}^{-2} .
\end{equation}

To avoid synchrotron self-Compton as the dominant cooling
mechanism requires $T_b\la 10^{12}$~K  (Kellermann \& Pauliny-Toth 1969). 
In addition, Readhead (1994) finds that equipartition between the
electron energy density and the magnetic energy density is obtained for
$T_b\sim 10^{11}$~K. 
Thus, the lower limits of $R_{\rm pc}^{\rm RS}\ga 0.1$ at $\nu_{\rm GHz}\sim 5$
in $L_{30}\sim 1$ objects imply $T_b$ that is just below 
the Readhead limit of $T_b\la 10^{11}$. 

When the source is optically thick, $T_b$ can be expressed in a simple
form, independent of $R$ and $L_{\nu}$, 
\begin{equation} 
T_b=\frac{c^2}{2\nu^2 k_{\rm B}}S_{\nu},
\end{equation}
which for $p=2$ gives
\begin{equation} 
T_b=2.96\times 10^{10}\nu_{\rm GHz}^{1/2}B^{-1/2} .
\end{equation}
Thus, equipartition between the electrons and the $B$ field is obtained
if the electrons emitting at $\nu$ are subject to 
$B\simeq 0.1\nu_{\rm GHz}$, and synchrotron self-Compton losses dominate
if $B< 10^{-3}\nu_{\rm GHz}$.

\subsubsection{Observational evidence for extended radio emission in AGN, ULXs, and
GBHs}

We conclude that the 5~GHz emission in AGN must be coming from a ``radio-sphere" which
is $\sim 100$ times larger than the 1~keV emitting region. In cool coronally active 
stars, interferometric mas scale
radio imaging reveal that transient emission originates in compact cores
associated with flare regions, but quiescent radio emission often originates 
on scales of a few stellar radii (G{\"u}del 2002), most likely from a population
of cooling electrons from previous flares, or possibly from electrons accelerated
locally on extended scales in shocks associated with coronal mass ejections 
(CMEs, Bastian et al. 1998; Bastian 2007).
If AGN coronae are magnetically dominated, then CMEs are likely to occur there as
well, as suggested by Merloni \& Fabian (2002, see also a related ``aborted jet"
mechanism by Ghisellini et al. 2004), and such CMEs may well be the source for the
extended radio emission in RQ AGN. Thus, similar mechanisms may be driving the
non-co spatial radio and X-ray emission in stars and AGN.

The extended pc scale radio emission seen in nearby RQ Seyferts
appears to originate in non-relativistic and relatively poorly collimated plasma
flows (e.g. Ulvestad et al. 1999; Middelberg et al. 2004), which may correspond to 
CMEs from the accretion disk. These outflows can be traced 
to kpc scale in nearby AGN (Gallimore et al. 2006), where they produce very low power,
and appear as slow buoyantly driven plasma which gradually dissipates and disappears.

It is interesting to note that in ULXs the radio emission also tends to be 
offset from the X-ray source, and is marginally resolved (Soria et al. 2006), 
indicating it is coming from a
more extended scale compared to the X-ray emission, and may thus also
be explained by emission from CMEs. However, the physical scale of a few tens
of pc implied by the offset (section 2.4) is much larger than predicted for
$R^{\rm RS}$, and thus if the coronal mechanisms is valid, higher quality 
observations should reveal that most of the radio flux originates from a compact 
core which overlaps with the X-ray position.

Spatially resolved jet-like emission in the radio is seen 
in Cygnus X-1 (Stirling et al. 2001) and in GRS 1915+105 (Dhawan et al. 2000). 
However, Heinz (2006) finds that the standard relativistic jet models for Cygnus X-1
appears to be inconsistent with the evidence for interaction with
the surrounding ISM (Gallo et al. 2005). One of the suggested solutions
is a much more massive outflow. It will be interesting to explore whether a non 
relativistic CME flow is a viable solution for Cygnus X-1.

Gallo et al. (2003) suggested that the radio emitting outflows in GBHs in the
low hard state are only mildly relativistic (cf. Heinz \& Merloni 2004). Clearly, 
a relatively slow and weakly collimated jet may be just an alternative name
for a CME, in particular when taken together with the suggestion that such a 
jet is fed by the accretion disk corona (Markoff et al. 2005). The identification
of the radio emission in GBHs with CMEs may allow us to gain some insight on
GBHs from studies of stellar coronae (e.g. on correlated radio and X-ray 
variability, Section 3.3.2).

\subsection{$B$ and $\gamma$}
Below we derive limits on the value of $B$ and $\gamma$ of the emitting electrons
within the radio-sphere based on the observed and predicted variability timescales.
We also provide constraints on these properties within the much more compact
X-ray emitting corona. 

\subsubsection{The radio-sphere}
Barvainis et al. (2005) and Anderson \& Ulvestad (2005) studied the variability
timescales at 8.5~GHz in luminous ($L_{30}\sim 1$) RQ PG quasars, and in
very low luminosity ($L_{30}\sim 10^{-5}$) AGN, respectively.
The shortest variability timescale in luminous AGN was found to be 
$t_{\rm var}\sim 10^7$~s,
and $\sim 10^5$~s in the very low luminosity AGN. 
These timescales provide upper limits on the size of the radio emission region,
$R<t_{\rm var}c$, based on the indications that the emission is not relativistically 
beamed (section 3.1).
Interestingly, these upper limits are within a factor of $\sim 2-5$ of 
the minimum size predicted for optically thick synchrotron emission 
(eq.22), suggesting that the synchrotron source is indeed close to being optically
thick. It also indicates that $t_{\rm var}$ could be
dominated by the light crossing time of the emitting regions, and thus the
intrinsic variability timescale could be shorter. What can we learn from 
the observed $t_{\rm var}$ on the electron cooling mechanism?
Below we briefly explore the following cooling mechanisms; synchrotron
emission, Compton scattering, Coulomb losses, and adiabatic expansion,
and the constraints they provide on $B$ and $\gamma$.

The time for an electron with an energy $\gamma m_ec^2$
to loose half its energy through synchrotron emission is 
\begin{equation} 
t_{\rm synch}=5.1\times 10^8\gamma^{-1}B^{-2}~s\ ,
\end{equation}
and the peak of the synchrotron emission occurs at 
\begin{equation} 
\nu_{\rm GHz}=\gamma^2B/820 
\end{equation}
(RL).
If indeed one observes at the peak frequency,
these two simple relations provide a lower limit on $B$ and an upper limit on
$\gamma$ from the observed cooling timescale, $t_{\rm var}$,
independently of any other properties of the emitting source, except the assumption that 
synchrotron cooling dominates other cooling processes.
Specifically, assuming $t_{\rm var}>t_{\rm synch}$ gives
\begin{equation} 
B> 6.8\times 10^4t_{\rm var}^{-2/3}\nu_{\rm GHz}^{-1/3}, 
\end{equation}  
and
\begin{equation} 
\gamma< 0.11 t_{\rm var}^{1/3}\nu_{\rm GHz}^{2/3}
\end{equation} 
(for $\gamma\gg 1$). 
Thus, for the luminous PG quasars, where $t_{\rm var}\sim 10^7$~s 
at 8.5~GHz, we find that the emission originates in electrons with
$\gamma < 100$, i.e $E < 50$~MeV, which reside in a $B > 0.7$~Gauss field. 
In very low luminosity AGN $t_{\rm var}\sim 10^5$~s, 
implying $\gamma < 21$ and $B > 15$~Gauss. 

The time for a relativistic electron to lose half its energy through Compton scattering 
is given by
\begin{equation} 
t_C=\frac{3cm}{4\sigma_TU_{\rm ph}\gamma} ,
\end{equation}
(RL) where $\sigma_T$ is the Thomson electron scattering
cross section, i.e.
\begin{equation} 
t_C=1.1\times 10^{10}\gamma^{-1}L_{46}^{-1}R_{\rm pc}^2~s\ .
\end{equation}
The requirement that $t_C\la t_{\rm var} \approx R/c$ 
implies that Compton cooling will be fast enough for electrons having
\begin{equation} 
\gamma\ga 110L_{46}^{-1}R_{\rm pc} .
\end{equation} 
In luminous $L_{46}\sim 1$ quasars, where the 8.5~GHz radio-sphere occurs at say
$R_{\rm pc}^{\rm RS}=0.055$ (eq.22), Compton cooling will be fast enough for
$\gamma\ga 6$, relevant for all synchrotron emitting electrons at 8.5~GHz (eq.30). 
However, in the lowest
luminosity AGN, say where $L_{46}\sim 10^{-5}$ and $R_{\rm pc}^{\rm RS}=1.7\times 10^{-4}$,
Compton cooling will be fast enough only for $\gamma\ga 1870$, much higher than the
likely $\gamma$ values of the 8.5~GHz emitting electrons (eq.30).
Thus, Compton cooling will be faster than the light crossing time only in 
luminous quasars.

The relative importance of Compton vs. synchrotron cooling can also be
estimated directly by comparing $U_{\rm ph}$ and $U_{\rm B}$.
In luminous quasars the requirement that $t_{\rm synch}<t_{\rm var}$ led
to $B\ga 0.7$ (see above), where synchrotron self-absorption implies
$R_{\rm pc}^{\rm RS}\ga 0.034$ (eq.19). For equipartition
$B_{\rm eq}=7.9$ (eq.21) at this radius, and thus Compton cooling can be
comparable or faster than the synchrotron cooling if $B\la B_{\rm eq}$.
This Compton cooling
will not be spectrally detectable.
Compton scattered photons peak at $\sim \gamma^2E_{\rm ph}$, which for
$\gamma\la 100$ and $E_{\rm ph}\sim 10-20$~eV (the UV bump peak) occurs at 
$E_{\rm ph}\la$100-200~keV. This luminosity of the X-ray Compton peak will 
be only of the order of the radio luminosity (as $t_{\rm synch}$ and $t_C$
are not dramatically different), which is lower by a few orders
of magnitude from the observed X-ray emission. 
In the lowest luminosity AGN ($L_{30}=10^{-5}$) the 8.5~GHz radio-sphere, 
for $B\ga 15$, is located at 
$R_{\rm pc}^{\rm RS}\ga 2.3\times 10^{-4}$, where $B_{\rm eq}=3.7$~Gauss, and 
since $B>B_{\rm eq}$ synchrotron cooling likely dominates.

Adiabatic cooling occurs on the sound crossing time, $t_{\rm ad}\sim R/c_s$, 
which must be longer than the light crossing time, $R/c$, and thus likely is 
too long to explain $t_{\rm var}$. Introducing clumping will reduce $t_{\rm ad}$
to $l/c_s$, where $l\ll R$ is the size of each clump.
However, the extent of the emitting region will now have to be
much larger than the size of a monolithic emitting region (eq.19), 
since the observed luminosity density provides a lower limit on the
total emitting surface area. If this area is comprised of small clumps
which cover only a small fraction of the total surface area, then the size
of the area must be correspondingly larger. Light travel time will then
lead to variability which is slower than $t_{\rm var}$ (which is within a factor
of few of $R^{\rm RS}/c$).

The energy loss timescale of relativistic electrons 
through elastic Coulomb collisions is
\begin{equation} 
t_{\rm coll}=2\times 10^{12}\gamma n^{-1}~s\ ,
\end{equation}
where $n$ is the ambient gas number density (Petrosian 1985). To obtain 
$t_{\rm coll}<t_{\rm var}$ requires $n\ga 10^7$~cm$^{-3}$ in luminous quasars,
and $n\ga 3\times 10^8$~cm$^{-3}$ in the lowest luminosity AGN. The implied
column in both cases is $\ga 10^{24}$~cm$^{-2}$, which becomes optically thick
to electron scattering, and therefore excluded as it would obscure the AGN. 
However, we cannot exclude such high densities if the synchrotron emission preferentially
occurs in high column dense gas off our line of sight (a.k.a. "the torus"). The ambient
gas will be heated by the Coulomb collisions with the relativistic electrons, but this 
heating is unlikely to observationally detectable as $L_R\ll L_X < L_{UV}$.

The free-free cooling time
\begin{equation} 
t_{\rm ff}=10^{17}(\ln \gamma)^{-1} n^{-1}~s,
\end{equation}
(Petrosian 1985) is $\sim 10^4$ times longer than $t_{\rm coll}$, and is therefore generally insignificant for cooling mildly relativistic electrons. 
Blundell \& Kuncic (2007) suggested that the radio emission of RQQs is entirely due to thermal free-free emission of hot ($T > 10^7$~K) plasma and not due to synchrotron. A basic difficulty we find with this conjecture is the prohibitively large free-free X-ray flux it involves.
The ratio of free free luminosity density at 1~keV and at 5~GHz in a $kT =$ 1~keV plasma is (e.g., eq. 5.14 in RL)

\begin{eqnarray}
\frac{L_{\nu _{\rm 1keV}} }{L_{\nu _{\rm 5GHz}}}=e^{-h(\nu _X - \nu _R) / kT}\frac{\bar{g}_{\rm ff}({\rm 1keV})}{\bar{g}_{\rm ff}({\rm 5GHz})} 
%\nonumber \\
%=e^{-h\nu _X / kT}\frac{\left( \frac{3kT}{\pi h\nu _X}\right)^{1/2}}{\frac{\sqrt{3}}{\pi }\ln \left( \frac{4kT}{\zeta h\nu _R}\right)}
%\nonumber \\
%\approx \frac{\sqrt{\pi}}{18}e^{-\frac{1keV}{kT}} \left( \frac{kT}{1keV}\right)^{1/2}
\nonumber \\
\approx 0.1e^{-\frac{{\rm 1keV}}{kT}} \left( \frac{kT}{{\rm 1keV}}\right)^{1/2}
\end{eqnarray}

\noindent where $\bar{g}_{\rm ff}$ are the gaunt factors and the weak logarithmic dependence of $\bar{g}_{\rm ff}({\rm 5GHz})$ on $kT/h\nu$ has been neglected. This ratio increases with $T$ and already overestimates by a few orders of magnitude the observed ratio of

\begin{equation}
\frac{L_{\nu _{\rm 1keV}}}{L_{\nu _{\rm 5GHz}}} = \frac{L_X/6.25\nu _{\rm 1keV}}{L_R/\nu _{\rm 5GHz}} 
= \frac{2\times10^5}{3\times10^8}=6.7\times 10^{-4}
\end{equation}

\noindent where we have used $L_X \approx 6.25 \nu  _{\rm 1keV} L_{\nu _{\rm 1keV}}$ (\S1) and $L_X / L_R \approx 2\times 10^5$ (eq. 1);
Not to mention the small contribution free-free emission must have to $L_X$ given the observed X-ray spectral slope of AGN, which is consistent with Comptonization.

To summarize, Compton cooling is effective only in luminous AGN, Coulomb losses are
effective only in dense gas, adiabatic losses are likely not fast enough, and free-free
cooling of the relativistic electrons is insignificant. Synchrotron cooling is a
plausible mechanism, which implies $B\ga 1-15$~Gauss for $t_{\rm var}\sim 10^5-10^7$~s
seen in AGN. 

An additional important conclusion from the fact that $t_{\rm var}\ga R^{\rm RS}/c$, 
is that the electrons must be accelerated locally, as the time it would take them to reach the
radio-sphere from the nucleus is likely to be much longer than the light crossing time.
Such local acceleration is observed in solar CMEs, presumably through the interaction
of a tangled magnetic field with the ambient medium (Bastian 2007), 
and a similar effect may be taking place in AGN.

\subsubsection{The X-ray corona}
As noted above, the $\sim 1-2~mm$ emission can come directly from the compact
coronal region, having $R\sim 0.01R^{\rm RS}$,
which produces the rapidly variable soft X-ray ($\sim 1$~keV) emission. 
Extrapolating 
$L_{\nu}$ assuming a flat spectral slope
($\alpha=0$) from 5~GHz to say 200~GHz, implies 
$\nu L_{\nu}/L_{\rm X}\sim 4\times 10^{-4}$, or a total radio to X-ray cooling ratio of 
$R/X\sim 10^{-3}$ from the coronal region. What constraints can be
obtained from this ratio on the coronal heating mechanisms? 

In the magnetically heated corona paradigm magnetic energy is converted to kinetic energy 
of fast electrons through reconnection, the fast electrons dissipate their energy and heat 
the corona, and the corona then cools by X-ray emission. 
In short, magnetic energy is converted to X-ray photons, and 
the average rate at which magnetic energy is annihilated must equal the X-ray luminosity
(Merloni \& Fabian 2001a). 
{\em What is the required minimum value for $B$?}
Assuming a uniform spherical shell with a radius $r$, within which reconnection propagates 
at a velocity $v_{\rm rec}$, leading to complete annihilation of $B$, yields a 
maximal luminosity of 

\begin{equation}
L_{\rm X}=B^2r^2v_{\rm rec}/2\ . 
\end{equation}

The equipartition field ($U_{\rm B}=U_{\rm ph}$) at a distance $R$ obeys
$L_{\rm bol}=B_{\rm eq}^2R^2c/2$. Combining both expressions gives 
$B/B_{\rm eq}\simeq R/r\sqrt{c/v_{\rm rec}}(L_X/L_{bol})^{1/2}$.
Since both $R/r$ and $c/v_{\rm rec}$ are likely $\gg 1$, and only a 
fraction of $B$ is annihilated, we get that in a magnetically heated
corona $B\gg B_{\rm eq}$. 

Can the power-law X-ray emission be produced directly by the fast electrons
through Compton scattering?  To produce the $R/X\sim 10^{-3}$ ratio 
requires $U_{\rm B}/U_{\rm ph}=10^{-3}$, however
above we concluded that $U_{\rm B}/U_{\rm ph}\gg 1$, and thus the observed X-ray emission cannot
originate from Compton scattering by the fast electrons which produce the mm radio emission. The X-ray emission then most likely originates from Compton cooling of 
the ambient non relativistic electrons in the corona, which is the commonly assumed
mechanism for the X-ray power-law emission (note that cooling of these background electrons
through low frequency cyclo-synchrotron 
emission is suppressed due to strong self-absorption, e.g. Ghisellini et al. 1998).

How do the fast electrons loose 99.9\% of their energy to the background gas? (to maintain
$R/X\sim 10^{-3}$). The remaining
mechanism is Coulomb collisions. The synchrotron/Coulomb cooling rate ratio is 
$4\times 10^3\gamma^2 B^2 n^{-1}$, and thus if the corona is non-homogeneous,
synchrotron losses will be largest for the
highest $\gamma$ electrons, at the regions with the highest $B$ and lowest $n$, and
Coulomb losses will dominate for the lower $\gamma$ electrons, in particular at regions
with high $n$ and low $B$. To obtain rough quantitative constraints on $n$ we assume a
steady state power-law electron energy distribution (eq.13), embedded in a homogeneous
corona with a uniform $B$ and $n$. The ratio of synchrotron to Coulomb cooling rates is

\begin{equation} 
\dot{E}_{\rm synch}=\int^{\gamma_1}_{\gamma_0}\frac{\gamma m_e c^2}{t_{\rm synch}}
\frac{dn}{d\gamma}d\gamma, \ \ \
\dot{E}_{\rm coll}=\int^{\gamma_1}_{\gamma_0}\frac{\gamma m_e c^2}{t_{\rm coll}}
\frac{dn}{d\gamma}d\gamma 
\end{equation}

which gives 

\begin{equation} 
\frac {\dot{E}_{\rm synch}}{\dot{E}_{\rm coll}}=4\times 10^3 \gamma_0^2B^2 n^{-1}
\frac{1-p}{3-p}\frac{(\gamma_1/\gamma_0)^{3-p}-1}{(\gamma_1/\gamma_0)^{1-p}-1} .
\end{equation}
For $\gamma_1/\gamma_0\gg 1$ and say $p=2$, we get
\begin{equation} 
\frac {\dot{E}_{\rm synch}}{\dot{E}_{\rm coll}}=4\times 10^3 \gamma_0\gamma_1B^2 n^{-1},
\end{equation}
which together with the requirement $\dot{E}_{\rm synch}/\dot{E}_{\rm coll}=10^{-3}$,
gives 
\begin{equation} 
n=4\times 10^6B^2\gamma_0\gamma_1~{\rm cm}^{-3}.
\end{equation}
For example, for plausible values of $B=100$~Gauss (e.g. $r^{-1}$ extrapolation from the 
radio-sphere in luminous AGN),
$\gamma_0=10, \gamma_1=100$, we get $n=4\times 10^{13}$~cm$^{-3}$, i.e rather high densities
which are likely to occur close to the surface of the accretion disk. In the above estimate we
assumed that the plasma is optically thin to the synchrotron radiation. As discussed above this is
true only above some threshold electron energy (which is a function of $B$). Inclusion of 
synchrotron self-absorption will reduce $\dot{E}_{\rm synch}$, and thus lower the implied
$n$.

Interestingly, the column heated by coulomb collisions is
\begin{equation} 
\Sigma_{\rm coll}=ct_{\rm coll}n=6\times 10^{22}\gamma
\end{equation}
or 
\begin{equation} 
\tau_{\rm es}=0.04\gamma .
\end{equation} 
Thus, for plausible $\gamma$ of a few to a few tens for the fast electrons, the collisionaly 
heated pathlength along the direction of $B$ has $\tau_{\rm es}\sim 0.1-1$, which would
lead to a significant Compton scattering optical depth for the photospheric disk emission 
(unless $B$ is close to parallel to the disk surface).
Compton scattering conserves the number of photons, and the observed number of 
photons in the X-rays is $\sim 10^{-3}$ of the UV bump photons (assuming the X-ray
luminosity is $\sim 10$\% of the UV luminosity, and the mean photon energy is
$\sim 100$ times larger). This implies that $C\times\tau_{\rm es}\sim 10^{-3}$, or a 
coronal covering factor $C\sim 10^{-2}-10^{-3}$ for $\tau_{\rm es}\sim 0.1-1$. Thus, the corona 
may be confined to small
``active" regions, presumably reconnecting coronal loops, as seen in solar flare 
activity (see also Haardt et al. 1994; Stern et al. 1995). 
The scarce covering of the accretion disk by the Comptonizing medium can also explain the 
vastly different amplitudes of X-ray and optical/UV variability and the complex (or lack of) connection between the two bands in some AGN (e.g., Maoz et al. 2002).
The slope of the Comptonized power-law 
emission is set by $\tau_{\rm es}$ and the electron temperature $T_e$. 
The observed slope, of order unity,
implies then $T_e\sim$ a few $10^9$~K (e.g. Eq. 7.45b in RL). 
{\em Will the corona cool mostly by Compton scattering?}
Another potentially significant cooling mechanism is thermal emission
(essentially pure free-free at this temperature), where the total cooling 
per unit volume is 
\begin{equation} 
\dot{E}_{\rm ff}=8\times 10^{-23}n^2T_9^{1/2}~{\rm erg~s}^{-1}~{\rm cm}^{-3} ,
\end{equation}
where $T_9=T_e/10^9$~K (Fig.7.1 in Dopita \& Sutherland 2003), implying an electron 
cooling time of 
\begin{equation} 
t'_{\rm ff}=\frac{3}{2}nkT_e/\dot{E}_{\rm ff}=2.6\times 10^{15}n^{-1}T_9^{1/2}~{\rm s} 
\end{equation}
(note that this is somewhat shorter than $t_{\rm ff}$ for relativistic electrons).
The Compton cooling time of the thermal electrons is estimated as follows. The mean energy lost
by a thermal electron upon scattering from a photon of energy $\epsilon = h\nu$ is
\begin{equation}
\Delta \epsilon=\frac{4kT_e}{m_ec^2}\epsilon\ ,
\end{equation} 
assuming $\epsilon\ll 4kT$ (RL). The total Compton cooling
per unit volume is then
\begin{equation}
\dot{E}_{\rm C}=n\sigma_{\rm es}\int \frac{F_{\nu}}{ch\nu}\frac{4kT_e}{m_ec^2}h\nu d\nu\ ,
\end{equation}
where $F_{\nu}$ is the incident (mono directional) flux density, or 
\begin{equation}
\dot{E}_{\rm C}=n\sigma_{\rm es}\frac{4kT_e}{m_ec^2}F ,
\end{equation}
where $F=\int F_{\nu}d\nu$. The Compton cooling time of the thermal electrons is then
\begin{equation}
t_C^{\rm th}=\frac{3}{2}nkT_e/\dot{E}_{\rm C}\ ,
\end{equation}
which gives the following simple expression
\begin{equation}
t_C^{\rm th}=\frac{3}{8}\frac{m_ec^2}{\sigma_{\rm es}F}=81.5T_5^{-4}~{\rm s}
\end{equation}
where $F=\sigma T_{\rm eff}^4$ is the underlying 
flux of the (assumed infinite slab) disk, and
$T_5=T_{\rm eff}/10^5$~K. 

Compton cooling dominates when $t_C^{\rm th}<t'_{\rm ff}$,
i.e. at 
\begin{equation} 
n<3\times 10^{13}T_5^4T_9^{1/2}~{\rm cm}^{-3}\ ,
\end{equation}
which is interestingly close to the lower limit on $n$ from the requirement that
the fast electrons dump 99.9\% of their energy as heat in the coronal gas.  

\subsubsection{A chromosphere?}
An additional potential implication, based on solar analogy, is the presence of a
transition ``chromospheric" layer between the corona and the UV emitting photosphere.
Such a layer can be heated by the smaller fraction of faster electrons 
(say $\gamma>100$), or ions, which deposit their energy at a larger depth, creating a deeper
and cooler layer below the coronal layer. Such a chromospheric layer, if at $T\sim 10^6$~K, 
will cool mostly by far UV and soft X-ray line emission (e.g. Sutherland \& Dopita 1993),
and it may be responsible for the ubiquitous soft excess emission feature
seen below $\sim 0.7$~keV, which is generally too strong to be explained
by reprocessing of the harder X-ray continuum (e.g. Gierli{\'n}ski \& Done 2004; 
Done \& Nayakshin 2007). We can estimate the chromospheric density $n_{\rm ch}$
by the following relation
\begin{equation} 
L_{\rm ch}=2\pi R_{\rm ch}^2C\Sigma_{\rm ch}n_{\rm ch}\Lambda 
\end{equation}
where the cooling function is $\Lambda\simeq 10^{-22}$~erg~s$^{-1}$~cm$^3$ at $T=10^6$~K 
(Sutherland \& Dopita 1993), the size of the X-ray coronal region in luminous AGN
$R_{\rm ch}\sim 3\times 10^{15}$~cm, the chromospheric column 
$\Sigma_{\rm ch}=6\times 10^{24}$~cm$^{-2}$ (for $\gamma=100$),
$C$ is the coronal covering fraction,
and $L_{\rm ch}=10^{44}$~erg~s$^{-1}$, assuming the soft X-ray feature carries 
$\sim 10$\% of the X-ray luminosity, which is $\sim 10$\% of the bolometric luminosity
in a luminous $L_{\rm bol}=10^{46}$~erg~s$^{-1}$ AGN. These values give 
$n_{\rm ch}\sim 3\times 10^{12}$~cm$^{-3}$, which is comparable to the coronal density 
inferred above (the apparent lack of a large density gradient between the corona and 
chromosphere is consistent with a disk which is hydrostatically balanced by radiation
pressure, rather than by a gas pressure gradient). The major difference in the microphysics 
of such a chromosphere, compared to earlier calculations of X-ray irradiated accretion
disks (e.g. Ross et al. 1999; Nayakshin et al. 2000), is that the chromosphere is 
collisionally heated by fast electrons, rather than photoionized by X-rays. This 
difference may lead to predictably different emission line ratios (e.g. Feldman et al.
2007), but it may be difficult to discern such differences as the chromospheric lines
are expected to be highly blended due to the large broadening by the Doppler 
effect in the inner disk.

\subsubsection{Correlated X-ray and mm variability}

Additional constraints on the typical $B$ and $\gamma$ within the corona can be 
obtained from measurements of the mm emission variability timescale, as estimated above
using the 5~GHz emission variability. In particular, for the 200~GHz emission in luminous AGN to
vary on a coronal light crossing time ($10^5$~s), requires $B>5.4$~Gauss. As discussed
above, the magnetically driven corona paradigm requires $B$ well above the radiation
equipartition value ($B_{\rm eq}\sim 300$~Gauss in luminous AGN). If the 
covering factor of the coronally active loops is small, then the required $B$ 
to generate the observed $L_{\rm X}$ will be higher, and may reach values
of $\sim 10^3-10^4$~Gauss seen in stellar coronally active regions.
This is also the expected value of $B$ if the coronal loops are 
driven by buoyancy out of the disk (as seen in the Sun) from regions at 
$T\simeq 10^5$~K dominated by radiation pressure.
The 200~GHz emitting electrons at $B=10^4$~Gauss have $\gamma\sim 4$, implying
a cooling time of $\sim 1$~s. Thus, the observed mm emission can vary rapidly, and
the observed variability will be dominated by the light crossing time effects over the
coronally active regions.

Current mm arrays are generally not sensitive enough to detect RQ AGN, which
are mostly mJy sources. However, more sensitive future mm arrays, in particular
ALMA, will be able to detect the predicted rapid (minutes to hours) mm variability 
at the sub mJy level in RQ AGN. Such studies, in particular with simultaneous X-ray
monitoring, can provide valuable insight on the coronal heating mechanism in AGN.

\section{Conclusions}
We find that RQ AGN lie on the G{\"u}del-Benz relation, $L_{\rm R}/L_{\rm X}=10^{-5}$,
found for coronally active stars. Since the X-ray emission in AGN most likely originates 
from a hot corona, and the corona
is likely to be magnetically heated, it is natural to associate the radio emission
in RQ AGN with a coronal origin as well. 

The ``coronal paradigm" for the radio emission implies a very compact source, which
is synchrotron self-absorbed at the GHz range. This implies that a compact flat spectrum 
source should generally be present in RQ AGN. Self-absorption will become negligible
at mm wavelengths, and this emission should originate from the same volume
producing the X-ray emission. The mm emission is likely to be strongly variable, 
and it should be correlated with the X-ray variability, possibly
as seen in the Neupert effect of stellar coronal flares.

Despite the factor of $\sim 10^{10}-10^{15}$ difference in $L_{\rm R}$ and $L_{\rm X}$ 
between AGN and coronally active stars, the estimated values of $B$ in the active regions 
of the two systems are comparable ($\sim 10^3-10^4$~Gauss), which may imply similar
underlying microphysics in both coronae. The physical mechanism underlying the
$L_{\rm R}/L_{\rm X}=10^{-5}$ ratio in these coronae remains a puzzle.

Applying the stellar analogy to the extended radio emission, we suggest that it originates 
from coronal mass ejections. This emission mechanism differrs from the alternative jet 
interpretation in that the outflow is relatively slow, and is not well confined, as 
suggested in observations of some nearby RQ Seyfert galaxies.

There are two additional populations of active radio and X-ray emitting objects, 
intermediate between stars and AGN. ULXs, where the four objects with radio and X-ray data also fall
close to the G{\"u}del-Benz relation, and overlap with the position of NGC~4395, the lowest
luminosity type 1 AGN. This is consistent with ULXs being scaled down AGN, rather than strongly
beamed GBHs. The other population is GBHs, where $L_{\rm R}$ is a factor of 10-100 lower
than expected from the G{\"u}del-Benz relation. This offset may result from a dependence
of $L_{\rm R}/L_{\rm X}$ on the local disk temperature. We show that
the Merloni et al. ``fundamental plane" relation, which incorporates both GBHs and AGN,
can be recast as a dependence of $L_{\rm R}/L_{\rm X}$ on the surface accretion disk 
effective temperature. This may indicate that the radio emission in GBHs has a
significant coronal component as well, in particular given the observed 
variability pattern, where $L_{\rm o}\propto -dL_{\rm X}/dt$, which is reminiscent of the
Neupert effect in stellar coronae.

We thank M. G{\"u}del for providing data in electronic form and A. Merloni for the fundamental-plane relation in his radio-quiet sub-sample.
We thank R. Antonucci and S. Jester for valuable comments on the manuscript.
We thank the referee for many useful comments that helped us improve the manuscript.
This research was supported in part by the Asher Space Research Institute at the Technion.

%\newpage

\newpage

\begin{table*}
 \centering
 \begin{minipage}{140mm}
  \caption{PG Quasar Data.}
  \begin{tabular}{@{}cccrrrrrr@{}}
  \hline
   Name     &  $z$    & M$_{\rm V}$ & Log R & $\alpha_{\rm ox}$ & Log $f_{\rm 1keV}$ &
Log L$_{\rm R}$ & Log L$_{\rm X}$ & C IV EW\\
   &  &  &  &  & \multicolumn{1}{c}{$\mu$Jy} & erg s$^{-1}$ & erg s$^{-1}$ & \multicolumn{1}{c}{\AA} \\
 \hline
0003+158   &  0.4505 &  $-$26.92 &    2.24 &   $-$1.38 &   $-$0.08 &  42.99 &  45.55  &  0  \\
0003+199   &  0.0260 &  $-$22.14 &   $-$0.57 &   $-$1.50 &    0.60 &  38.36 &  43.80  &  0  \\
0007+106   &  0.0893 &  $-$23.85 &    2.29 &   $-$1.43 &    0.20 &  41.93 &  44.49  &  0.8  \\
0026+129   &  0.1452 &  $-$24.71 &    0.03 &   $-$1.50 &    0.02 &  39.97 &  44.70  &  0  \\
0043+039   &  0.3859 &  $-$26.16 &   $-$0.92 &   $<-$2.00 &   $<-$1.45 &  39.44 &  $<$44.05  &  22.3  \\
0049+171   &  0.0643 &  $-$21.81 &   $-$0.49 &   $-$1.27 &    0.04 &  38.27 &  44.05  &  -  \\
0050+124   &  0.0587 &  $-$23.77 &   $-$0.48 &   $-$1.56 &    0.21 &  38.97 &  44.18  &  0.4  \\
0052+251   &  0.1544 &  $-$24.64 &   $-$0.62 &   $-$1.37 &    0.17 &  39.35 &  44.93  &  0  \\
0157+001   &  0.1632 &  $-$24.61 &    0.33 &   $-$1.60 &   $-$0.57 &  40.13 &  44.23  &  -  \\
0804+761   &  0.1005 &  $-$24.44 &   $-$0.22 &   $-$1.52 &    0.39 &  39.88 &  44.78  &  -  \\
0838+770   &  0.1318 &  $-$23.83 &   $<-$0.96 &   $-$1.54 &   $-$0.46 &  $<$38.59 &  44.16  &  -  \\
0844+349   &  0.0644 &  $-$23.31 &   $-$1.52 &   $-$1.54 &   $-$0.01 &  37.81 &  43.99  &  0.6  \\
0921+525   &  0.0352 &  $-$21.25 &    0.17 &   $-$1.41 &    0.03 &  38.62 &  43.52  &  -  \\
0923+129   &  0.0287 &  $-$21.59 &    0.32 &   $-$1.41 &    0.37 &  38.94 &  43.69  &  -  \\
0923+201   &  0.1929 &  $-$24.56 &   $-$0.85 &   $-$1.57 &   $-$0.54 &  39.16 &  44.38  &  0.8  \\
0934+013   &  0.0505 &  $-$21.43 &   $-$0.42 &   $-$1.39 &   $-$0.07 &  38.13 &  43.73  &  -  \\
0947+396   &  0.2059 &  $-$24.21 &   $-$0.60 &   $-$1.33 &   $-$0.28 &  39.07 &  44.72  &  0  \\
0953+414   &  0.2341 &  $-$25.65 &   $-$0.36 &   $-$1.50 &   $-$0.13 &  39.98 &  44.98  &  0.1  \\
1001+054   &  0.1610 &  $-$24.07 &   $-$0.30 &   $-$2.13 &   $-$2.05 &  39.38 &  42.74  &  11.8  \\
1004+130   &  0.2404 &  $-$25.97 &    2.36 &   $-$1.82 &   $-$0.63 &  42.70 &  44.48  &  16.6  \\
1011$-$040   &  0.0584 &  $-$22.70 &   $-$1.00 &   $-$2.01 &   $-$1.32 &  38.09 &  42.60  &  1.0  \\
1012+008   &  0.1865 &  $-$24.79 &   $-$0.30 &   $-$1.66 &   $-$0.85 &  39.50 &  44.06  &  -  \\
1022+519   &  0.0449 &  $-$21.40 &   $-$0.64 &   $-$1.34 &   $-$0.04 &  37.75 &  43.66  &  -  \\
1048+342   &  0.1667 &  $-$24.02 &   $<-$1.00 &   $-$1.52 &   $-$0.60 &  $<$38.61 &  44.22  &  -  \\
1048$-$090   &  0.3461 &  $-$25.83 &    2.58 &   $-$1.41 &   $-$0.26 &  42.97 &  45.16  &  -  \\
1049$-$005   &  0.3596 &  $-$25.93 &   $-$0.60 &   $-$1.56 &   $-$0.67 &  39.81 &  44.78  &  0.3  \\
1100+772   &  0.3115 &  $-$25.86 &    2.51 &   $-$1.39 &   $-$0.10 &  42.92 &  45.24  &  0.4  \\
1103$-$006   &  0.4232 &  $-$25.96 &    2.43 &   $-$1.51 &   $-$0.68 &  42.88 &  44.90  &  0  \\
1114+445   &  0.1438 &  $-$24.01 &   $-$0.89 &   $-$1.62 &   $-$0.83 &  38.65 &  43.87  &  4.0  \\
1115+407   &  0.1542 &  $-$23.74 &   $-$0.77 &   $-$1.45 &   $-$0.43 &  38.65 &  44.33  &  0.3  \\
1116+215   &  0.1765 &  $-$25.57 &   $-$0.14 &   $-$1.57 &   $-$0.04 &  40.20 &  44.83  &  0  \\
1119+120   &  0.0500 &  $-$22.49 &   $-$0.82 &   $-$1.58 &   $-$0.30 &  38.05 &  43.48  &  -  \\
1121+422   &  0.2248 &  $-$24.38 &   $<-$1.00 &   $-$1.59 &   $-$1.00 &  $<$38.68 &  44.10  &  0  \\
1126$-$041   &  0.0601 &  $-$23.00 &   $-$0.77 &   $-$2.13 &   $-$1.63 &  38.44 &  42.32  &  5.8  \\
1149$-$110   &  0.0489 &  $-$21.90 &   $-$0.06 &   $-$1.42 &    0.05 &  38.80 &  43.83  &  -  \\
1151+117   &  0.1759 &  $-$24.07 &   $<-$1.15 &   $-$1.46 &   $-$0.44 &  $<$38.54 &  44.43  &  -  \\
1202+281   &  0.1654 &  $-$23.75 &   $-$0.72 &   $-$1.27 &   $-$0.11 &  38.74 &  44.70  &  0  \\
1211+143   &  0.0810 &  $-$24.60 &    0.20 &   $-$1.57 &    0.29 &  40.13 &  44.54  &  0  \\
1216+069   &  0.3318 &  $-$26.33 &    0.22 &   $-$1.44 &   $-$0.29 &  40.62 &  45.10  &  0  \\
1226+023   &  0.1575 &  $-$27.15 &    3.06 &   $-$1.47 &    0.90 &  43.97 &  45.67  &  0  \\
1229+204   &  0.0640 &  $-$23.06 &   $-$0.96 &   $-$1.49 &   $-$0.01 &  38.29 &  44.00  &  -  \\
1244+026   &  0.0480 &  $-$21.77 &   $-$0.28 &   $-$1.60 &   $-$0.18 &  38.62 &  43.58  &  -  \\
1259+593   &  0.4770 &  $-$26.82 &   $<-$1.00 &   $-$1.75 &   $-$1.13 &  $<$39.70 &  44.53  &  0  \\
1302$-$102   &  0.2783 &  $-$26.60 &    2.27 &   $-$1.58 &   $-$0.17 &  42.99 &  45.09  &  0  \\
1307+085   &  0.1545 &  $-$24.56 &   $-$1.00 &   $-$1.52 &   $-$0.16 &  38.94 &  44.59  &  0  \\
1309+355   &  0.1825 &  $-$24.76 &    1.26 &   $-$1.71 &   $-$1.01 &  41.05 &  43.89  &  2.2  \\
1310$-$108   &  0.0343 &  $-$21.35 &   $-$1.00 &   $-$1.52 &   $-$0.19 &  37.63 &  43.30  &  -  \\
1322+659   &  0.1676 &  $-$24.23 &   $-$0.92 &   $-$1.40 &   $-$0.14 &  38.80 &  44.68  &  0  \\
1341+258   &  0.0864 &  $-$22.71 &   $<-$0.92 &   $-$1.53 &   $-$0.44 &  $<$38.25 &  43.83  &  -  \\
1351+236   &  0.0553 &  $-$22.40 &   $-$0.59 &   $-$1.52 &   $-$0.77 &  37.93 &  43.11  &  -  \\
1351+640   &  0.0880 &  $-$24.08 &    0.64 &   $-$1.78 &   $-$0.74 &  40.29 &  43.53  &  5.3  \\
1352+183   &  0.1508 &  $-$24.13 &   $-$0.96 &   $-$1.50 &   $-$0.39 &  38.76 &  44.38  &  0  \\
1354+213   &  0.3011 &  $-$24.34 &   $<-$1.10 &   $-$1.39 &   $-$0.71 &  $<$38.64 &  44.60  &  -  \\
1402+261   &  0.1643 &  $-$24.48 &   $-$0.64 &   $-$1.58 &   $-$0.36 &  39.27 &  44.44  &  0.7  \\
1404+226   &  0.0978 &  $-$22.93 &   $-$0.33 &   $-$1.55 &   $-$0.60 &  38.67 &  43.77  &  1.5  \\
1411+442   &  0.0897 &  $-$23.54 &   $-$0.89 &   $-$2.03 &   $-$1.56 &  38.50 &  42.72  &  10.3  \\
1415+451   &  0.1133 &  $-$23.41 &   $-$0.77 &   $-$1.51 &   $-$0.51 &  38.50 &  43.99  &  0  \\
1416$-$129   &  0.1292 &  $-$24.55 &    0.06 &   $-$1.56 &    0.05 &  40.27 &  44.65  &  0  \\
1425+267   &  0.3635 &  $-$26.18 &    1.73 &   $-$1.63 &   $-$1.25 &  41.82 &  44.21  &  2.1  \\
1426+015   &  0.0863 &  $-$24.05 &   $-$0.55 &   $-$1.46 &    0.32 &  39.30 &  44.58  &  -  \\
1427+480   &  0.2203 &  $-$24.04 &   $<-$0.80 &   $-$1.52 &   $-$0.75 &  $<$38.88 &  44.30  &  0  \\
1435$-$067   &  0.1288 &  $-$24.55 &   $-$1.15 &   $-$1.63 &   $-$0.32 &  38.79 &  44.28  &  -  \\
\hline
\end{tabular}
\end{minipage}
\end{table*}

\newpage

\setcounter{table}{0}
\begin{table*}
 \centering
 \begin{minipage}{140mm}
  \caption{--- Continued}
  \begin{tabular}{@{}cccrrrrrr@{}}
  \hline
   Name     &  $z$    & M$_{\rm V}$ & Log R & $\alpha_{\rm ox}$ & Log $f_{\rm 1keV}$ &
Log L$_{\rm R}$ & Log L$_{\rm X}$ & EW C IV \\
   &  &  &  &  & \multicolumn{1}{c}{$\mu$Jy} & erg s$^{-1}$ & erg s$^{-1}$ & \multicolumn{1}{c}{\AA} \\
 \hline
1440+356   &  0.0777 &  $-$23.49 &   $-$0.43 &   $-$1.38 &    0.37 &  38.94 &  44.54  &  0  \\
1444+407   &  0.2676 &  $-$25.18 &   $<-$1.10 &   $-$1.57 &   $-$0.66 &  $<$38.99 &  44.55  &  0  \\
1448+273   &  0.0648 &  $-$23.30 &   $-$0.60 &   $-$1.59 &   $-$0.16 &  38.58 &  43.86  &  -  \\
1501+106   &  0.0365 &  $-$22.76 &   $-$0.44 &   $-$1.64 &    0.14 &  38.92 &  43.65  &  -  \\
1512+370   &  0.3713 &  $-$25.93 &    2.28 &   $-$1.43 &   $-$0.45 &  42.63 &  45.03  &  0  \\
1519+226   &  0.1357 &  $-$23.76 &   $-$0.05 &   $-$1.51 &   $-$0.41 &  39.47 &  44.24  &  -  \\
1534+580   &  0.0305 &  $-$21.44 &   $-$0.15 &   $-$1.38 &    0.19 &  38.38 &  43.54  &  0  \\
1535+547   &  0.0389 &  $-$22.15 &   $-$0.85 &   $<-$2.17 &   $<-$1.89 &  37.84 &  $<$41.76  &  3.9  \\
1543+489   &  0.4009 &  $-$25.60 &   $-$0.82 &   $-$1.67 &   $-$1.13 &  39.44 &  44.40  &  0  \\
1545+210   &  0.2643 &  $-$25.63 &    2.62 &   $-$1.38 &   $-$0.07 &  42.91 &  45.14  &  0  \\
1552+085   &  0.1191 &  $-$23.72 &   $-$0.35 &   $-$1.77 &   $-$0.98 &  39.10 &  43.55  &  -  \\
1612+261   &  0.1308 &  $-$23.77 &    0.45 &   $-$1.41 &   $-$0.08 &  40.02 &  44.53  &  0  \\
1613+658   &  0.1291 &  $-$24.22 &    0.00 &   $-$1.21 &    0.29 &  39.55 &  44.89  &  -  \\
1617+175   &  0.1137 &  $-$23.95 &   $-$0.14 &   $-$1.64 &   $-$0.34 &  39.76 &  44.15  &  -  \\
1626+554   &  0.1317 &  $-$23.54 &   $-$0.96 &   $-$1.37 &   $-$0.15 &  38.52 &  44.48  &  0  \\
1700+518   &  0.2892 &  $-$26.44 &    0.37 &   $<-$2.29 &   $<-$2.39 &  40.75 &  $<$42.90  &  94  \\
1704+608   &  0.3721 &  $-$26.38 &    2.81 &   $-$1.62 &   $-$0.83 &  43.29 &  44.64  &  2.4  \\
2112+059   &  0.4597 &  $-$27.26 &   $-$0.49 &   $-$2.11 &   $-$1.79 &  40.51 &  43.87  &  26.7  \\
2130+099   &  0.0631 &  $-$23.23 &   $-$0.49 &   $-$1.47 &    0.39 &  39.00 &  44.35  &  0.6  \\
2209+184   &  0.0697 &  $-$23.14 &    2.15 &   $-$1.49 &   $-$0.14 &  41.45 &  43.94  &  -  \\
2214+139   &  0.0657 &  $-$23.39 &   $-$1.30 &   $-$2.02 &   $-$1.41 &  38.03 &  42.63  &  1.1  \\
2233+134   &  0.3263 &  $-$25.18 &   $-$0.55 &   $-$1.66 &   $-$0.95 &  39.65 &  44.42  &  -  \\
2251+113   &  0.3255 &  $-$26.24 &    2.56 &   $-$1.86 &   $-$1.28 &  43.06 &  44.09  &  3.5  \\
2304+042   &  0.0426 &  $-$21.58 &   $-$0.60 &   $-$1.29 &    0.24 &  38.14 &  43.89  &  -  \\
2308+098   &  0.4336 &  $-$26.24 &    2.27 &   $-$1.35 &   $-$0.24 &  42.76 &  45.35  &  0.2  \\
\hline
\end{tabular}
\end{minipage}
\end{table*}


\begin{thebibliography}{}

\bibitem[Anderson et al.(2004)]{2004ApJ...603...42A} Anderson, J.~M., 
Ulvestad, J.~S., \& Ho, L.~C.\ 2004, ApJ, 603, 42 

\bibitem[Anderson \& Ulvestad(2005)]{2005ApJ...627..674A} Anderson, J.~M., 
\& Ulvestad, J.~S.\ 2005, ApJ, 627, 674 

\bibitem[Balbus \& Hawley(1998)]{1998RvMP...70....1B} Balbus, S.~A., \& 
Hawley, J.~F.\ 1998, Reviews of Modern Physics, 70, 1 

\bibitem[Ballantyne et al.(2001)]{2001MNRAS.327...10B} Ballantyne, D.~R., 
Ross, R.~R., \& Fabian, A.~C.\ 2001, MNRAS, 327, 10 

\bibitem[Barvainis et al.(1996)]{1996AJ....111.1431B} Barvainis, R., 
Lonsdale, C., \& Antonucci, R.\ 1996, AJ, 111, 1431 

\bibitem[Barvainis et al.(2005)]{2005ApJ...618..108B} Barvainis, R., 
Leh{\'a}r, J., Birkinshaw, M., Falcke, H., \& Blundell, K.~M.\ 2005, ApJ, 
618, 108 

\bibitem[Baskin \& Laor(2005)]{2005MNRAS.356.1029B} Baskin, A., \& Laor, 
A.\ 2005, MNRAS, 356, 1029 

\bibitem[Bastian et al.(1998)]{1998ARA&A..36..131B} Bastian, T.~S., Benz, 
A.~O., \& Gary, D.~E.\ 1998, ARA\&A, 36, 131 

\bibitem[Bastian(2007)]{2007arXiv0704.3108B} Bastian, T.~S.\ 2007, ArXiv 
e-prints, 704, arXiv:0704.3108 

\bibitem[Begelman et al.(1984)]{1984RvMP...56..255B} Begelman, M.~C., 
Blandford, R.~D., \& Rees, M.~J.\ 1984, Reviews of Modern Physics, 56, 255 

\bibitem[Benz \& G{\"u}del(1994)]{1994A&A...285..621B} Benz, A.~O., \& G{\"u}del, 
M.\ 1994, A\&A, 285, 621 

\bibitem[Blackman \& Field(2000)]{2000MNRAS.318..724B} Blackman, E.~G., \& 
Field, G.~B.\ 2000, MNRAS, 318, 724

\bibitem[Blundell et al.(1996)]{1996ApJ...468L..91B} Blundell, K.~M., 
Beasley, A.~J., Lacy, M., \& Garrington, S.~T.\ 1996, ApJ, 468, L91 

\bibitem[Blundell \& Beasley(1998)]{1998MNRAS.299..165B} Blundell, K.~M., 
\& Beasley, A.~J.\ 1998, MNRAS, 299, 165 

\bibitem[Blundell et al.(2003)]{2003ApJ...591L.103B} Blundell, K.~M., 
Beasley, A.~J., \& Bicknell, G.~V.\ 2003, ApJ, 591, L103 

\bibitem[Blundell \& Kuncic(2007)]{}
Blundell, M.~B. \& Kuncic, Z.\ 2007, ApJ, 668, L103

\bibitem[Boroson \& Green(1992)]{1992ApJS...80..109B} Boroson, T.~A., \& 
Green, R.~F.\ 1992, ApJS, 80, 109 (BG92) 

\bibitem[Brandt et al.(2000)]{2000ApJ...528..637B} Brandt, W.~N., Laor, A., 
\& Wills, B.~J.\ 2000, ApJ, 528, 637

\bibitem[Brinkmann et al.(2000)]{2000A&A...356..445B} Brinkmann, W., 
Laurent-Muehleisen, S.~A., Voges, W., Siebert, J., Becker, R.~H., 
Brotherton, M.~S., White, R.~L., \& Gregg, M.~D.\ 2000, A\&A, 356, 445 

\bibitem[Brocksopp et al.(2006)]{2006MNRAS.366..953B} Brocksopp, C., 
Starling, R.~L.~C., Schady, P., Mason, K.~O., Romero-Colmenero, E., \& 
Puchnarewicz, E.~M.\ 2006, MNRAS, 366, 953 

\bibitem[Capetti \& Balmaverde(2007)]{2007A&A...469...75C} Capetti, A., \& 
Balmaverde, B.\ 2007, A\&A, 469, 75 

\bibitem[Condon et al.(1991)]{1991ApJ...378...65C} Condon, J.~J., Huang, 
Z.-P., Yin, Q.~F., \& Thuan, T.~X.\ 1991, ApJ, 378, 65 

\bibitem[Corbel et 
al.(2000)]{2000A&A...359..251C} Corbel, S., Fender, R.~P., Tzioumis, A.~K., 
Nowak, M., McIntyre, V., Durouchoux, P., \& Sood, R.\ 2000, A\&A, 359, 251 

\bibitem[Crenshaw et al.(1999)]{1999ApJ...516..750C} Crenshaw, D.~M., 
Kraemer, S.~B., Boggess, A., Maran, S.~P., Mushotzky, R.~F., \& Wu, C.-C.\ 
1999, ApJ, 516, 750 

\bibitem[Crenshaw et al.(2001)]{2001ApJ...555..633C} Crenshaw, D.~M., 
Kraemer, S.~B., Bruhweiler, F.~C., \& Ruiz, J.~R.\ 2001, ApJ, 555, 633 

\bibitem[Crenshaw et al.(2004)]{2004ApJ...612..152C} Crenshaw, D.~M., 
Kraemer, S.~B., Gabel, J.~R., Schmitt, H.~R., Filippenko, A.~V., Ho, L.~C., 
Shields, J.~C., \& Turner, T.~J.\ 2004, ApJ, 612, 152 

\bibitem[Czerny et al.(2004)]{2004A&A...420....1C} Czerny, B., 
R{\'o}{\.z}a{\'n}ska, A., Dov{\v c}iak, M., Karas, V., \& Dumont, A.-M.\ 
2004, A\&A, 420, 1 

\bibitem[de Bruyn(1976)]{1976A&A....52..439D} de Bruyn, A.~G.\ 1976, A\& A, 52, 439 

\bibitem[Dewangan et al.(2004)]{2004ApJ...608L..57D} Dewangan, G.~C., 
Miyaji, T., Griffiths, R.~E., \& Lehmann, I.\ 2004, ApJ, 608, L57 

\bibitem[Dhawan et al.(2000)]{2000ApJ...543..373D} Dhawan, V., Mirabel, 
I.~F., \& Rodr{\'{\i}}guez, L.~F.\ 2000, ApJ, 543, 373 

\bibitem[di Matteo et al.(1997)]{1997MNRAS.291..805D} di Matteo, T., 
Celotti, A., \& Fabian, A.~C.\ 1997, MNRAS, 291, 805 

\bibitem[di Matteo et al.(1999)]{1999MNRAS.304..809D} di Matteo, T., 
Celotti, A., \& Fabian, A.~C.\ 1999, MNRAS, 304, 809 

\bibitem[Done \& Nayakshin(2007)]{2007MNRAS.377L..59D} Done, C., \& 
Nayakshin, S.\ 2007, MNRAS, 377, L59 

\bibitem[Done et 
al.(2007)]{2007A&ARv..15....1D} Done, C., Gierli{\'n}ski, M., \& Kubota, A.\ 2007, 
A\&A Review, 15, 1

\bibitem[Dopita \& Sutherland(2003)]{2003adu..book.....D} Dopita, M.~A., \& 
Sutherland, R.~S.\ 2003, Astrophysics of the diffuse universe, Berlin, New 
York: Springer, 2003.~Astronomy and astrophysics library, ISBN 3540433627

\bibitem[Dove et al.(1997)]{1997ApJ...487..747D} Dove, J.~B., Wilms, J., \& 
Begelman, M.~C.\ 1997, ApJ, 487, 747

\bibitem[Draine \& Li(2007)]{2007ApJ...657..810D} Draine, B.~T., \& Li, A.\ 
2007, ApJ, 657, 810

\bibitem[Eardley \& Lightman(1975)]{1975ApJ...200..187E} Eardley, D.~M., \& 
Lightman, A.~P.\ 1975, ApJ, 200, 187

\bibitem[Elvis et al.(1994)]{1994ApJS...95....1E} Elvis, M., et al.\ 1994, 
ApJS, 95, 1 

\bibitem[Evans \& Koratkar(2004)]{2004ApJS..150...73E} Evans, I.~N., \& 
Koratkar, A.~P.\ 2004, ApJS, 150, 73 

\bibitem[Falcke et al.(1996)]{1996ApJ...471..106F} Falcke, H., Sherwood, 
W., \& Patnaik, A.~R.\ 1996, ApJ, 471, 106

\bibitem[Falcke et al.(2004)]{2004A&A...414..895F} Falcke, H., K{\"o}rding, 
E., \& Markoff, S.\ 2004, A\&A, 414, 895 

\bibitem[Feldman et al.(2007)]{2007ApJ...660.1674F} Feldman, U., Landi, E., 
\& Doschek, G.~A.\ 2007, ApJ, 660, 1674 

\bibitem[Field \& Rogers(1993)]{1993ApJ...403...94F} Field, G.~B., \& 
Rogers, R.~D.\ 1993, ApJ, 403, 94 

\bibitem[Gabel et al.(2003)]{2003ApJ...583...178G} Gabel et al. 2003, ApJ, 583, 178 

\bibitem[Galeev et al.(1979)]{1979ApJ...229..318G} Galeev, A.~A., Rosner, 
R., \& Vaiana, G.~S.\ 1979, ApJ, 229, 318 

\bibitem[Gallimore et al.(2006)]{2006AJ....132..546G} Gallimore, J.~F., 
Axon, D.~J., O'Dea, C.~P., Baum, S.~A., \& Pedlar, A.\ 2006, AJ, 132, 546 

\bibitem[Gallo et al.(2003)]{2003MNRAS.344...60G} Gallo, E., Fender, R.~P., 
\& Pooley, G.~G.\ 2003, MNRAS, 344, 60 

\bibitem[Gallo et al.(2005)]{2005Natur.436..819G} Gallo, E., Fender, R., 
Kaiser, C., Russell, D., Morganti, R., Oosterloo, T., \& Heinz, S.\ 2005, 
Nature, 436, 819 

\bibitem[Gallo et al.(2006)]{2006MNRAS.370.1351G} Gallo, E., Fender, R.~P., 
Miller-Jones, J.~C.~A., Merloni, A., Jonker, P.~G., Heinz, S., Maccarone, 
T.~J., \& van der Klis, M.\ 2006, MNRAS, 370, 1351 

\bibitem[George et al.(2000)]{2000ApJ...531...52G} George, I.~M., Turner, 
T.~J., Yaqoob, T., Netzer, H., Laor, A., Mushotzky, R.~F., Nandra, K., \& 
Takahashi, T.\ 2000, ApJ, 531, 52 

\bibitem[Ghisellini et al.(1998)]{1998MNRAS.297..348G} Ghisellini, G., 
Haardt, F., \& Svensson, R.\ 1998, MNRAS, 297, 348 

\bibitem[Ghisellini et al.(2004)]{2004A&A...413..535G} Ghisellini, G., 
Haardt, F., \& Matt, G.\ 2004, A\&A, 413, 535 

\bibitem[Gierli{\'n}ski \& Done(2004)]{2004MNRAS.349L...7G} Gierli{\'n}ski, 
M., \& Done, C.\ 2004, MNRAS, 349, L7 

\bibitem[Gilli et al.(2007)]{2007A&A...463...79G} Gilli, R., Comastri, A., 
\& Hasinger, G.\ 2007, A\&A, 463, 79 

\bibitem[Ginzburg \& Syrovatskii(1969)]{1969ARA&A...7..375G}
Ginzburg V. L. \& Syrovatskii S. I.\ 1969, ARA\&A, 7, 375 

\bibitem[G{\"u}del(2002)]{2002AR..40..217G} G{\"u}del, M.\ 2002, ARAA, 
40, 217	

\bibitem[G{\"u}del(2004)]{2004A&ARv..12...71G} G{\"u}del, M.\ 2004, A\&A Review, 
12, 71 

\bibitem[G{\"u}del \& Benz(1993)]{1993ApJ...405L..63G} G{\"u}del, M., \& Benz, 
A.~O.\ 1993, ApJ, 405, L63 

\bibitem[Haardt \& Maraschi(1993)]{1993ApJ...413..507H} Haardt, F., \& 
Maraschi, L.\ 1993, ApJ, 413, 507

\bibitem[Haardt et al.(1994)]{1994ApJ...432L..95H} Haardt, F., Maraschi, 
L., \& Ghisellini, G.\ 1994, ApJ, 432, L95 

\bibitem[Haas et al.(2000)]{2000A&A...354..453H} Haas, M., M{\"u}ller, 
S.~A.~H., Chini, R., Meisenheimer, K., Klaas, U., Lemke, D., Kreysa, E., \& 
Camenzind, M.\ 2000, A\&A, 354, 453 

\bibitem[H(2006)]{2006ApJ} Hardcastle, M. J., Croston, J. H., \& Kraft, R. P.\ 
2007, ApJ, in press, ArXiv Astrophysics e-prints, arXiv:0707.2865

\bibitem[Heinz(2006)]{2006ApJ...636..316H} Heinz, S.\ 2006, ApJ, 636, 316 

\bibitem[Heinz \& Merloni(2004)]{2004MNRAS.355L...1H} Heinz, S., \& Merloni, 
A.\ 2004, MNRAS, 355, L1 

\bibitem[Ho et al.(1995)]{1995ApJS...98..477H} Ho, L.~C., Filippenko, 
A.~V., \& Sargent, W.~L.\ 1995, ApJS, 98, 477 

\bibitem[Ho et al.(1997)]{1997ApJS..112..315H} Ho, L.~C., Filippenko, 
A.~V., \& Sargent, W.~L.~W.\ 1997, ApJS, 112, 315 
	
\bibitem[Ho \& Ulvestad(2001)]{2001ApJS..133...77H} Ho, L.~C., \& Ulvestad, 
J.~S.\ 2001, ApJS, 133, 77 

\bibitem[Hummel et al.(1987)]{1987A&AS...70..517H} Hummel, E., van der 
Hulst, J.~M., Keel, W.~C., \& Kennicutt, R.~C., Jr.\ 1987, A\&A Supplemet, 70, 517 

\bibitem[Iwasawa et al.(2000)]{2000MNRAS.318..879I} Iwasawa, K., Fabian, 
A.~C., Almaini, O., Lira, P., Lawrence, A., Hayashida, K., \& Inoue, H.\ 
2000, MNRAS, 318, 879 

\bibitem[Jester et al.(2005)]{2005AJ....130..873J} Jester, S., et al.\ 
2005, AJ, 130, 873 

\bibitem[Just et al.(2007)]{2007ApJ...665.1004J} Just, D. W.,  Brandt, W. N., Shemmer, O., Steffen, A. T.,  
Schneider, D. P., Chartas, G., \& Garmire, G. P.\ 
2007, ApJ, 665, 1004

\bibitem[Kaaret et al.(2003)]{2003Sci...299..365K} Kaaret, P., Corbel, S., 
Prestwich, A.~H., \& Zezas, A.\ 2003, Science, 299, 365 

\bibitem[Kaaret et al.(2006)]{2006ApJ...646..174K} Kaaret, P., Simet, 
M.~G., \& Lang, C.~C.\ 2006, ApJ, 646, 174 

\bibitem[Kaspi et al.(2004)]{2004AJ....127.2631K} Kaspi, S., Brandt, W.~N., 
Collinge, M.~J., Elvis, M., \& Reynolds, C.~S.\ 2004, AJ, 127, 2631 
1999, ApJ, 516, 750 

\bibitem[Kellermann \& Pauliny-Toth(1969)]{1969ApJ...155L..71K} Kellermann, 
K.~I., \& Pauliny-Toth, I.~I.~K.\ 1969, ApJ, 155, L71 

\bibitem[Kellermann et al.(1989)]{1989AJ.....98.1195K} Kellermann, K.~I., 
Sramek, R., Schmidt, M., Shaffer, D.~B., \& Green, R.\ 1989, AJ, 98, 1195 

\bibitem[Kellermann et al.(1994)]{1994AJ....108.1163K} Kellermann, K.~I., 
Sramek, R.~A., Schmidt, M., Green, R.~F., \& Shaffer, D.~B.\ 1994, AJ, 
108, 1163 

\bibitem[K{\"o}rding et al.(2005)]{2005A&A...436..427K} K{\"o}rding, E., 
Colbert, E., \& Falcke, H.\ 2005, A\&A, 436, 427 

\bibitem[K{\"o}rding et al.(2006)]{2006MNRAS.372.1366K} K{\"o}rding, E.~G., 
Jester, S., \& Fender, R.\ 2006, MNRAS, 372, 1366 

\bibitem[Kukula et al.(1995)]{1995MNRAS.276.1262K} Kukula, M.~J., Pedlar, 
A., Baum, S.~A., \& O'Dea, C.~P.\ 1995, MNRAS, 276, 1262 

\bibitem[Kukula et al.(1998)]{1998MNRAS.297..366K} Kukula, M.~J., Dunlop, 
J.~S., Hughes, D.~H., \& Rawlings, S.\ 1998, MNRAS, 297, 366

\bibitem[Kriss et al.(1980)]{1980ApJ...242..492K} Kriss, G.~A., Canizares, 
C.~R., \& Ricker, G.~R.\ 1980, ApJ, 242, 492 

\bibitem[Laor \& Brandt(2002)]{2002ApJ...569..641L} Laor, A., \& Brandt, 
W.~N.\ 2002, ApJ, 569, 641 

\bibitem[Leipski et al.(2006)]{2006A&A...455..161L} Leipski, C., Falcke, 
H., Bennert, N.,  Huettemeister, S.\ 2006, A\&A, 455, 161 

\bibitem[Malzac et al.(2003)]{2003A&A...407..335M} Malzac, J., Belloni, T., 
Spruit, H.~C., \& Kanbach, G.\ 2003, A\&A, 407, 335 

\bibitem[Malzac et al.(2004)]{2004MNRAS.351..253M} Malzac, J., Merloni, A., 
\& Fabian, A.~C.\ 2004, MNRAS, 351, 253 

\bibitem[Maoz et al.(2002)]{2002AJ....124.1988M}
Maoz, D., Markowitz, A., Edelson, R., Nandra, K.\ 2002, AJ, 124, 1988

\bibitem[Maoz(2007)]{2007MNRAS.377.1696M} Maoz, D.\ 2007, MNRAS, 377, 1696 

\bibitem[Maoz et al.(1998)]{1998AJ....116...55M} Maoz, D., Koratkar, A., 
Shields, J.~C., Ho, L.~C., Filippenko, A.~V., \& Sternberg, A.\ 1998, AJ, 
116, 55

\bibitem[Markoff et al.(2005)]{2005ApJ...635.1203M} Markoff, S., Nowak, 
M.~A., \& Wilms, J.\ 2005, ApJ, 635, 1203 

\bibitem[Marscher(1977)]{1977ApJ...216..244M} Marscher, A.~P.\ 1977, ApJ, 
216, 244 

\bibitem[Matt et al.(1997)]{1997MNRAS.289..175M} Matt, G., Fabian, A.~C., 
\& Reynolds, C.~S.\ 1997, MNRAS, 289, 175 

\bibitem[Merloni et al.(2000)]{2000MNRAS.318L..15M} Merloni, A., Di Matteo, 
T., \& Fabian, A.~C.\ 2000, MNRAS, 318, L15 

\bibitem[Merloni \& Fabian(2001)]{2001MNRAS.321..549M} Merloni, A., \& 
Fabian, A.~C.\ 2001a, MNRAS, 321, 549 

\bibitem[Merloni \& Fabian(2001)]{2001MNRAS.328..958M} Merloni, A., \& 
Fabian, A.~C.\ 2001b, MNRAS, 328, 958 

\bibitem[Merloni \& Fabian(2002)]{2002MNRAS.332..165M} Merloni, A., \& 
Fabian, A.~C.\ 2002, MNRAS, 332, 165

\bibitem[Merloni et al.(2003)]{2003MNRAS.345.1057M} Merloni, A., Heinz, S., 
\& di Matteo, T.\ 2003, MNRAS, 345, 1057 

\bibitem[Middelberg et al.(2004)]{2004A&A...417..925M} Middelberg, E., et 
al.\ 2004, A\&A, 417, 925 

\bibitem[Miller et al.(1990)]{1990MNRAS.244..207M} Miller, L., Peacock, 
J.~A., \& Mead, A.~R.~G.\ 1990, MNRAS, 244, 207 

\bibitem[Miller et al.(1993)]{1993MNRAS.263..425M} Miller, P., Rawlings, 
S., \& Saunders, R.\ 1993, MNRAS, 263, 425 

\bibitem[Miller(2005)]{2005Ap&SS.300..227M} Miller, J.~M.\ 2005, Astrophysics
\& Space Science, 
300, 227

\bibitem[Miller et al.(2006)]{2006ApJ...653..525M} Miller, J.~M., Homan, 
J., Steeghs, D., Rupen, M., Hunstead, R.~W., Wijnands, R., Charles, P.~A., 
\& Fabian, A.~C.\ 2006a, ApJ, 653, 525 

\bibitem[Miller et al.(2006)]{2006ApJ...652..163M} Miller, B.~P., Brandt, 
W.~N., Gallagher, S.~C., Laor, A., Wills, B.~J., Garmire, G.~P., \& 
Schneider, D.~P.\ 2006b, ApJ, 652, 163 

\bibitem[Moran et al.(2005)]{2005AJ....129.2108M} Moran, E.~C., Eracleous, 
M., Leighly, K.~M., Chartas, G., Filippenko, A.~V., Ho, L.~C., \& Blanco, 
P.~R.\ 2005, AJ, 129, 2108 

\bibitem[Mundell et al.(1995)]{1995MNRAS.275...67M} Mundell, C.~G., 
Holloway, A.~J., Pedlar, A., Meaburn, J., Kukula, M.~J., \& Axon, D.~J.\ 
1995, MNRAS, 275, 67 

\bibitem[Mushotzky et al.(1993)]{1993ARA&A..31..717M} Mushotzky, R.~F., 
Done, C., \& Pounds, K.~A.\ 1993, ARAA, 31, 717 

\bibitem[Mushotzky(2006)]{2006AdSpR..38.2793M} Mushotzky, R.\ 2006, 
Advances in Space Research, 38, 2793 

\bibitem[Nagar et al.(2002)]{2002A&A...392...53N} Nagar, N.~M., Falcke, H., 
Wilson, A.~S., \& Ulvestad, J.~S.\ 2002, A\&A, 392, 53 

\bibitem[Narayan 
\& McClintock(2008)]{2008NewAR..51..733N} Narayan, R., \& McClintock, J.~E.
\ 2008, New Astronomy Review, 51, 733 

\bibitem[Nayakshin et al.(2000)]{2000ApJ...537..833N} Nayakshin, S., 
Kazanas, D., \& Kallman, T.~R.\ 2000, ApJ, 537, 833

\bibitem[Neupert(1968)]{1968ApJ...153L..59N} Neupert, W.~M.\ 1968, ApJ, 
153, L59 

\bibitem[Nied{\'z}wiecki(2005)]{2005MNRAS.356..913N} Nied{\'z}wiecki, A.\ 
2005, MNRAS, 356, 913 

\bibitem[Panessa et al.(2006)]{2006A&A...455..173P} Panessa, F., Bassani, 
L., Cappi, M., Dadina, M., Barcons, X., Carrera, F.~J., Ho, L.~C., \& 
Iwasawa, K.\ 2006, A\&A, 455, 173 

\bibitem[Panessa et al.(2007)]{2007A&A...467..519P} Panessa, F., Barcons, 
X., Bassani, L., Cappi, M., Carrera, F.~J., Ho, L.~C., \& Pellegrini, S.\ 
2007, A\&A, 467, 519

\bibitem[Petrosian(1985)]{1985ApJ...299..987P} Petrosian, V.\ 1985, ApJ, 
299, 987 

\bibitem[Polletta et al.(2000)]{2000A&A...362...75P} Polletta, M., 
Courvoisier, T.~J.-L., Hooper, E.~J., \& Wilkes, B.~J.\ 2000, A\&A, 362, 75 

\bibitem[Ponti et al.(2006)]{2006MNRAS.368..903P} Ponti, G., Miniutti, G., 
Cappi, M., Maraschi, L., Fabian, A.~C., \& Iwasawa, K.\ 2006, MNRAS, 368, 
903 

\bibitem[Pounds et al.(2004)]{2004MNRAS.350...10P} Pounds, K.~A., Reeves, 
J.~N., King, A.~R., \& Page, K.~L.\ 2004, MNRAS, 350, 10 

\bibitem[Poutanen \& Fabian(1999)]{1999MNRAS.306L..31P} Poutanen, J., \& 
Fabian, A.~C.\ 1999, MNRAS, 306, L31 

\bibitem[Readhead(1994)]{1994ApJ...426...51R} Readhead, A.~C.~S.\ 1994, 
ApJ, 426, 51

\bibitem[Ross \& Fabian(1993)]{1993MNRAS.261...74R} Ross, R.~R., \& Fabian, 
A.~C.\ 1993, MNRAS, 261, 74 

\bibitem[Ross et al.(1999)]{1999MNRAS.306..461R} Ross, R.~R., Fabian, 
A.~C., \& Young, A.~J.\ 1999, MNRAS, 306, 461 

\bibitem[Rybicki \& Lightman(1979)]{1979rpa..book.....R} Rybicki, G.~B., \&
Lightman, A.~P.\ 1979, Radiative Processes in Astrophysics,
New York, Wiley-Interscience, 1979 (RL)

\bibitem[Rykoff et al.(2007)]{2007astro.ph..3497R} Rykoff, E.~S., Miller, 
J.~M., Steeghs, D., \& Torres, M.~A.~P.\ 2007, ArXiv Astrophysics e-prints, 
arXiv:astro-ph/0703497

\bibitem[Sahayanathan \& Misra(2005)]{2005ApJ...628..611S} Sahayanathan, 
S., \& Misra, R.\ 2005, ApJ, 628, 611 

\bibitem[Salvato et al.(2004)]{2004ApJ...600L..31S} Salvato, M., Greiner, 
J., \& Kuhlbrodt, B.\ 2004, ApJ, 600, L31 

\bibitem[Sanders et al.(1989)]{1989ApJ...347...29S} Sanders, D.~B., 
Phinney, E.~S., Neugebauer, G., Soifer, B.~T., \& Matthews, K.\ 1989, ApJ, 
347, 29

\bibitem[Scelsi et al.(2007)]{2007A&A...473..589S}
Scelsi, L., Maggio, A., Micela, G., Briggs, K., \& G{\"u}del, M.\ 2007, 
A\&A, 473, 589

\bibitem[Schmidt \& Green(1983)]{1983ApJ...269..352S} Schmidt, M., \& 
Green, R.~F.\ 1983, ApJ, 269, 352 

\bibitem[Shakura \& Syunyaev(1973)]{1973A&A....24..337S} Shakura, N.~I., \& 
Syunyaev, R.~A.\ 1973, A\&A, 24, 337 

\bibitem[Shapiro et al.(1976)]{1976ApJ...204..187S} Shapiro, S.~L., 
Lightman, A.~P., \& Eardley, D.~M.\ 1976, ApJ, 204, 187

\bibitem[Sikora et al.(2007)]{2007ApJ...658..815S} Sikora, M., Stawarz, 
{\L}., \& Lasota, J.-P.\ 2007, ApJ, 658, 815 

\bibitem[Sopp \& Alexander(1991)]{1991MNRAS.251P..14S} Sopp, H.~M., \& 
Alexander, P.\ 1991, MNRAS, 251, 14P 

\bibitem[Soria et al.(2006)]{2006MNRAS.370.1666S} Soria, R., Kuncic, Z., 
Broderick, J.~W., \& Ryder, S.~D.\ 2006, MNRAS, 370, 1666 

\bibitem[Steffen et al.(2006)]{2006AJ....131.2826S} Steffen, A.~T., Strateva,
I., Brandt, W. N., Alexander, D.~M., Koekemoer, A. M., Lehmer, B. D., 
Schneider, D.~P. \& Vignali, C.\ 2006, AJ, 131, 2826

\bibitem[Stern et al.(1995)]{1995ApJ...449L..13S} Stern, B.~E., Poutanen, 
J., Svensson, R., Sikora, M., \& Begelman, M.~C.\ 1995, ApJ, 449, L13

\bibitem[Stirling et al.(2001)]{2001MNRAS.327.1273S} Stirling, A.~M., 
Spencer, R.~E., de la Force, C.~J., Garrett, M.~A., Fender, R.~P., \& 
Ogley, R.~N.\ 2001, MNRAS, 327, 1273 

\bibitem[Sunyaev \& Titarchuk(1980)]{1980A&A....86..121S} Sunyaev, R.~A., 
\& Titarchuk, L.~G.\ 1980, A\&A, 86, 121 

\bibitem[Sutherland \& Dopita(1993)]{1993ApJS...88..253S} Sutherland, 
R.~S., \& Dopita, M.~A.\ 1993, ApJS, 88, 253 

\bibitem[Terashima \& Wilson(2003)]{2003ApJ...583..145T} Terashima, Y., \& 
Wilson, A.~S.\ 2003, ApJ, 583, 145 

\bibitem[Terlevich et al.(1992)]{1992MNRAS.255..713T} Terlevich, R., 
Tenorio-Tagle, G., Franco, J., \& Melnick, J.\ 1992, MNRAS, 255, 713 

\bibitem[Thim et al.(2004)]{2004AJ....127.2322T} Thim, F., Hoessel, J.~G., 
Saha, A., Claver, J., Dolphin, A., \& Tammann, G.~A.\ 2004, AJ, 127, 2322 

\bibitem[Ulvestad \& Wilson(1984)]{1984ApJ...278..544U} 
Ulvestad, J. S. \& Wilson, A. S.\ 1984, ApJ, 278, 544

\bibitem[Ulvestad \& Ho(2001)]{2001ApJ...562L.133U} Ulvestad, J.~S., \& Ho, 
L.~C.\ 2001, ApJ, 562, L133 

\bibitem[Ulvestad et al.(1998)]{1998ApJ...496..196U} Ulvestad, J.~S., Roy, 
A.~L., Colbert, E.~J.~M., \& Wilson, A.~S.\ 1998, ApJ, 496, 196 

\bibitem[Ulvestad et al.(1999)]{1999ApJ...517L..81U} Ulvestad, J.~S., 
Wrobel, J.~M., Roy, A.~L., Wilson, A.~S., Falcke, H., \& Krichbaum, T.~P.\ 
1999, ApJ, 517, L81 

\bibitem[Ulvestad et al.(2005)]{2005ApJ...621..123U} Ulvestad, J.~S., 
Antonucci, R.~R.~J., \& Barvainis, R.\ 2005, ApJ, 621, 123 

\bibitem[Wang et al.(2006)]{2006ApJ...645..890W} Wang, R., Wu, X.-B., \& 
Kong, M.-Z.\ 2006, ApJ, 645, 890 

\bibitem[White et al.(2007)]{2007ApJ...654...99W} White, R.~L., Helfand, 
D.~J., Becker, R.~H., Glikman, E., \& de Vries, W.\ 2007, ApJ, 654, 99 

\bibitem[Wilson \& Willis(1980)]{1980ApJ...240..429W} 
Wilson, A. S. \& Willis, A. G.\ 1980, ApJ, 240, 429

\bibitem[Zheng et al.(1997)]{1997ApJ...475..469Z} Zheng, W., Kriss, G.~A., 
Telfer, R.~C., Grimes, J.~P., \& Davidsen, A.~F.\ 1997, ApJ, 475, 469 

\end{thebibliography}
\end{document}